\documentclass[sigconf]{acmart}

\AtBeginDocument{%
  }

\setcopyright{acmlicensed}
\copyrightyear{2026}
\acmYear{2026}
\acmDOI{XXXXXXX.XXXXXXX}

\usepackage[super]{nth}
\usepackage{amsthm,amsmath}
\usepackage{appendix}
\usepackage{float}
\usepackage{bm}
\usepackage{dsfont}
\usepackage{graphicx}
\usepackage{caption}
\usepackage{subfigure}
\usepackage{subcaption} 
\usepackage{float}
\usepackage{enumitem}
\usepackage{array}
\usepackage{makecell}
\usepackage{multirow}
\usepackage{tabularx}
\usepackage{soul}
\usepackage{url}
\usepackage[flushleft]{threeparttable}
\usepackage[ruled,vlined,linesnumbered]{algorithm2e}
\usepackage{mathtools}
\DeclareMathOperator{\Rep}{Rep}
\usepackage{dsfont} 
\usepackage{booktabs,makecell,array}
\newcolumntype{L}[1]{>{\raggedright\arraybackslash}p{#1}}

\usepackage{xcolor}
\definecolor{stdgray}{gray}{0.48}
\newcommand{\mstd}[2]{\ensuremath{#1\,{\color{stdgray}\scriptstyle\pm\,#2}}}

\theoremstyle{definition}
\newtheorem{definition}{Definition}
\usepackage[noend]{algpseudocode} 

\usepackage{amssymb}

\newcommand{\pval}[1]{{\scriptsize\textcolor{black}{#1}}}

\begin{document}

\title[Adverse Online Social Interactions: A Multi-Level Evolutionary Analysis]{
Adverse Online Social Interactions: A Multi-Level Evolutionary Analysis of Local Patterns, Diffusion, and Community Disruption}

\author{Xueqi Cheng}
\affiliation{%
  \institution{Florida State University}
  \city{Tallahassee}
  \country{USA}
}
\email{xc25@fsu.edu}
\authornote{Equal contribution and co-first authors.}

\author{Qinwen Ge}
\authornotemark[1]
\affiliation{%
  \institution{Vanderbilt University}
  \city{Nashville}
  \country{USA}
}
\email{qinwen.ge@vanderbilt.edu}

\author{Hamid	Karimi}
\affiliation{%
  \institution{Utah State University}
  \city{Logan}
  \country{USA}
}
\email{hamid.karimi@usu.edu}

\author{Yushun	Dong}
\affiliation{%
  \institution{Florida State University}
  \city{Tallahassee}
  \country{USA}
}
\email{yd24f@fsu.edu}

\author{Tyler Derr}
\affiliation{%
  \institution{Vanderbilt University}
  \city{Nashville}
  \country{USA}
}
\email{tyler.derr@vanderbilt.edu}

\renewcommand{\shortauthors}{Cheng et al.}

\begin{abstract}
Adverse social interactions (ASIs) can shape how online communities evolve over the time. However, structural-based ASIs and content-based ASIs are often studied separately and at a single analytical scale. In this study, we propose a multi-level framework to examine how adverse social interactions appear locally, spread through neighborhoods, and disrupt cohesive subgroups. Using large-scale datasets from X and Bluesky, we analyze friend and foe patterns at the micro level, peer influence through matched triadic designs at the meso level, and subgroup disruption against random and recommendation-based references at the macro level. Our results show that structural disconnection and toxic communication provide complementary signals: structural negativity more persistently marks subgroup disruption, while toxic communication captures broader conflict both within and across communities. These findings suggest that adverse social interactions are multi-scale processes that influence how online communities form, fracture, and evolve. Our source code is publicly available at \url{https://github.com/XueqiC/Adverse-Social-Interactions}.
\end{abstract}



\maketitle

\vspace{-1.25ex}
\section{Introduction}

Adverse social interactions (ASI) such as blocking~\cite{tong2017efficient, pham2020multi,dey2017centrality}, unfollowing~\cite{wu2020mining, kwak2011fragile, kivran2011impact}, social disconnection~\cite{twenge2013does, jorge2019social}, and toxic posting~\cite{sheth2022defining, ali2021understanding} have become increasingly prevalent across online social platforms. While blocking and unfollowing are not inherently expressive of hostility and may be individually rational in some contexts, user intent is typically unobserved in large-scale data. Accordingly, we do not infer motivation from these actions. Recent studies reveal that these behaviors not only affect individual experiences but also reshape the structural and information diffusion of social networks over time~\cite{sonnemans2006dynamics, kivran2011impact, cote2019evolution, cheng2025bts}. Understanding such cumulative network-level evolutionary impact of ASI is therefore essential for developing theoretical models of social network dynamics and informing the design and governance of online communities~\cite{kwak2017understanding, offer2021negative, shen2014evolution, ganley2009ties, cheng2025edge}.

Despite their importance, the evolutionary and network-level effects of ASI remain insufficiently understood, limiting our ability to explain observed changes in social cohesion, user engagement, and polarization over time~\cite{harrigan2020negative, candellone2025negative, marlowe2017digital}. Most existing studies examine ASI from either a structural perspective that focuses on relational actions such as blocking, unfollowing, and antagonistic ties, or a content-based perspective that focuses on toxic or uncivil expressions, and these two facets are often studied separately~\cite{doreian2009partitioning,diaz2025mathematical,zigron2019help,ryberg2008networked,seckin2025identifying,humprecht2020hostile,scheuerman2021framework, cheng2025edgeclass}. Building on these two research directions, existing work has also analyzed ASI using different lenses across multiple scales~\cite{sasahara2021social, leskovec2010predicting, bolibar2016macro,liu2022weak}. At the micro-level, prior work has analyzed signed networks to model antagonistic links and local balance, examined user roles in toxic exchanges, and identified behavioral patterns around unpopular or “outsider” nodes~\cite{felmlee2016toxic,budak2011limiting,belaza2019social}. At the meso-level, studies have explored emotional contagion and triadic closure, demonstrating how negative actions, such as insults or blocks, can cascade to others through social influence~\cite{esmailian2014mesoscopic,huang2018will,song2019triadic,li2023information}. At the macro-level, computational and social science literature has shown that structural-based negativity in both connection patterns and content can contribute to echo chambers, increased affective polarization, and large-scale network fragmentation~\cite{cinelli2021echo,pedersen2019analyzing,candellone2025negative}. However, several key research gaps remain. First, most prior studies analyze either structural-based negativity, such as blocking and unfollowing, or content-level toxicity in isolation, 
without directly comparing 
the two~\cite{xu2013structures,kaiser2022partisan,saveski2021structure,kumar2023understanding}. Second, few works examine how negative behavior evolves across different levels, from individual interactions to network-wide outcomes~\cite{myers2014bursty,snijders2017modeling}. Third, few studies compare ASI dynamics across platforms with different interaction structures, leaving the role of platform context in ASI’s evolutionary impact unclear~\cite{goel2023hatemongers,israeli2022going}.

To mitigate these gaps, we study two definitions of ASI that involve structural-based disconnection, 
such as blocking and unfollowing, with content-based toxicity derived from user posts; this enables a consistent analysis of both relational and communicative hostility. Specifically, we focus on the evolutionary impact of ASI by analyzing the dynamics of negative behaviors and how they affect social network structure, i.e., 
our research is focus on answering:

\vspace{0.75ex}
\parbox{0.9\linewidth}{
\textit{How do structural- and content-based adverse social interactions exist and evolve locally, spread through neighborhoods, and disrupt online social communities?}}

\vspace{0.75ex}

\noindent To enable this, we collect and curate two large-scale real-world datasets, drawn respectively from X (formerly Twitter) and Bluesky.

\begin{figure}[t]
  \centering
  \includegraphics[width=0.98\columnwidth]{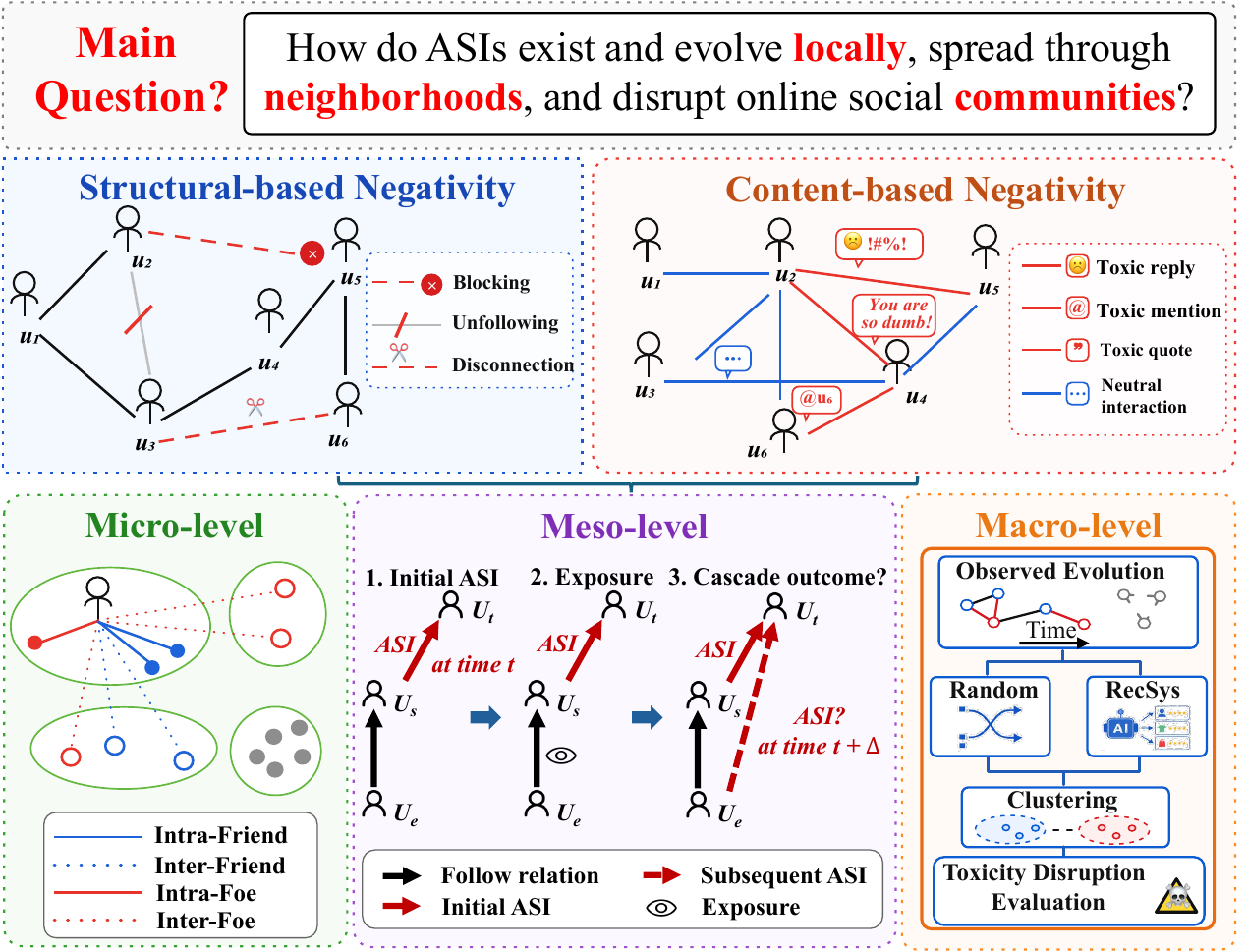}
  \vspace{-2ex}
  \caption{Overview of our multi-scale ASI analysis framework, covering structural and content-based negativity. The micro analysis examines the existence and evolution of ASI locally through friend/foe patterns, the meso analysis estimates how ASI spreads through neighborhoods, and the macro analysis evaluates the disruption impact of ASI.}
  \label{fig:overall}
  \vskip -2.5ex
\end{figure}

Building on these definitions and datasets, we instantiate this research question through the three-level analysis framework summarized in Figure~\ref{fig:overall}. Specifically, at the micro level, we examine how negative behaviors are distributed within and across communities and how these patterns vary across user prominence groups. At the meso level, we quantify the causal peer influence of ASI by estimating whether one user’s adverse social interaction increases the subsequent ASI propensity of their immediate neighbors, thereby characterizing local diffusion and cascade formation. At the macro level, we assess how ASI-related tie dissolution and formation accumulate into cohesive subgroup disruption, and compare the observed changes against random and recommendation-based counterfactual mechanisms. Together, these analyses support a comparative study of how distinct platform contexts shape the dynamics and impact of ASI. Our contributions are summarized as follows:
\begin{itemize}[left=0pt]
    \item We introduce a unified formulation of adverse social interactions that captures two complementary channels: structural-based disconnection, such as blocking, unfollowing and disconnections, and content-based toxicity from user posts.
    \item We develop a three-level analysis framework to study the evolutionary impact of ASI, covering local friend/foe patterns at the micro level, matched triadic influence at the meso level, and cohesive subgroup disruption at the macro level.
    \item We apply the framework to two large-scale datasets from X and Bluesky, enabling a cross-platform comparison of how different interaction structures shape the emergence, diffusion, and community-level impact of ASI.
\end{itemize}

In the following sections, We first review related work in Section~\ref{sec:related}, then formalize ASI definitions and describe the X and Bluesky datasets in Section~\ref{sec:data}. After that we present our three-level analysis of ASI's evolutionary impact from Section~\ref{sec:micro} to ~\ref{sec:macro}. After these, we discuss the findings, limitations, and future work in Section~\ref{sec:discuss}-~\ref{sec:limitations}.

\section{Related Work}\label{sec:related}

Existing studies examine ASI in social networks from two primary angles: structural disconnection and content-based toxicity. Structural work focuses on modeling unfollowing, blocking, and antagonistic ties using signed networks and balance theory, revealing that ASI often concentrates between communities and helps delineate social boundaries~\cite{doreian2009partitioning,diaz2025mathematical,zigron2019help,ryberg2008networked}. Other studies highlight how users who frequently block or are unfriended occupy marginalized or polarized positions~\cite{baysha2020dividing,kaiser2022partisan,zhu2022political}. Separately, research on content-level toxicity has developed methods for detecting hostile or uncivil language and has shown that exposure to toxic comments can escalate further aggression or disengagement~\cite{seckin2025identifying,humprecht2020hostile,scheuerman2021framework}. However, these two lines of work are often studied separately: research on structural-based negativity rarely considers post-level content, while studies of toxic content seldom account for the surrounding and evolving online social network structures.

Prior work also tends to focus on a single analytical level. Micro-level studies examine local motifs or community patterns~\cite{felmlee2016toxic,budak2011limiting,belaza2019social}; meso-level work studies triadic closure and contagion through friends-of-friends~\cite{esmailian2014mesoscopic,huang2018will,song2019triadic,li2023information}; and macro-level studies analyze polarization, echo chambers, and fragmentation~\cite{cinelli2021echo,pedersen2019analyzing,candellone2025negative}. Few works provide a temporally grounded multi-level view linking individual negative actions to broader structural outcomes.

\begin{figure*}[h!]
    \centering

    \begin{minipage}[t]{0.8\textwidth}
        \centering
        \includegraphics[width=\linewidth]{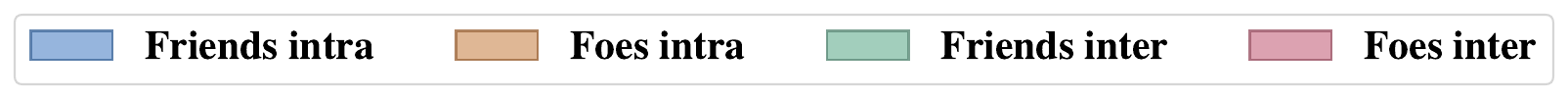}
    \end{minipage}

    \begin{minipage}[t]{0.48\textwidth}
        \centering
        \includegraphics[width=\linewidth]{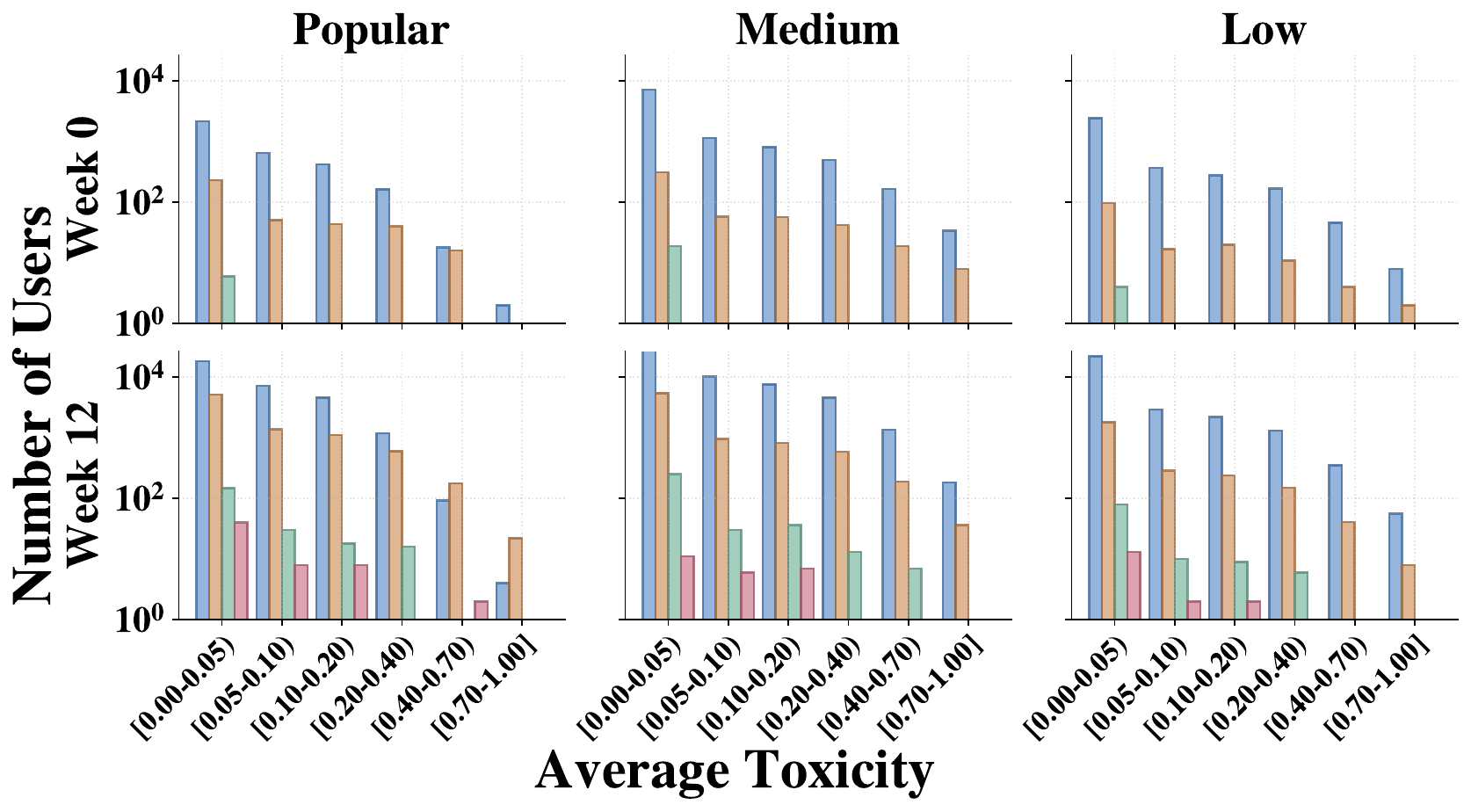}
        \smallskip
        \textbf{(a) X, Definition I}
    \end{minipage}
    \hfill
    \begin{minipage}[t]{0.48\textwidth}
        \centering
        \includegraphics[width=\linewidth]{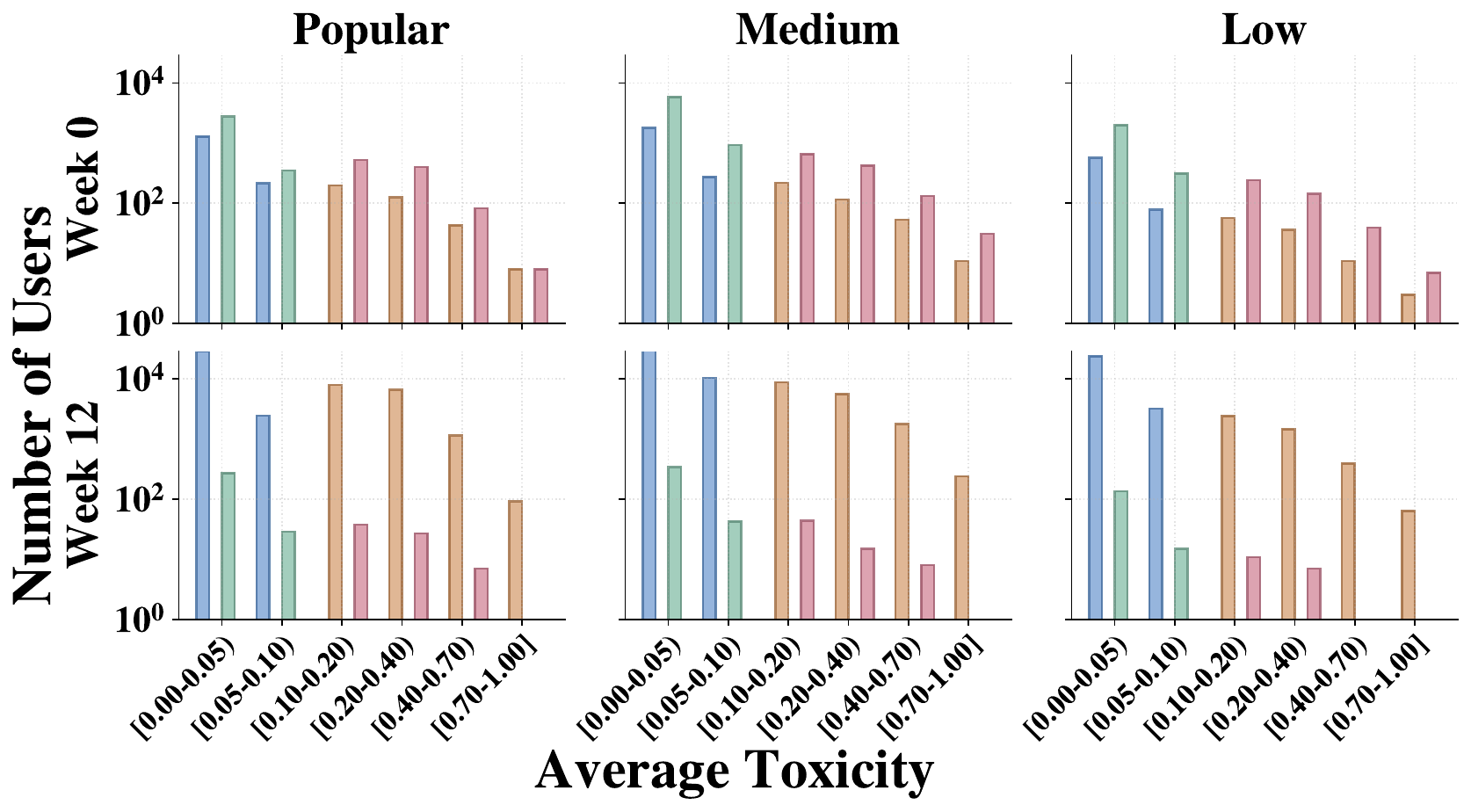}
        \smallskip
        \textbf{(b) X, Definition II}
    \end{minipage}

    \begin{minipage}[t]{0.48\textwidth}
        \centering
        \includegraphics[width=\linewidth]{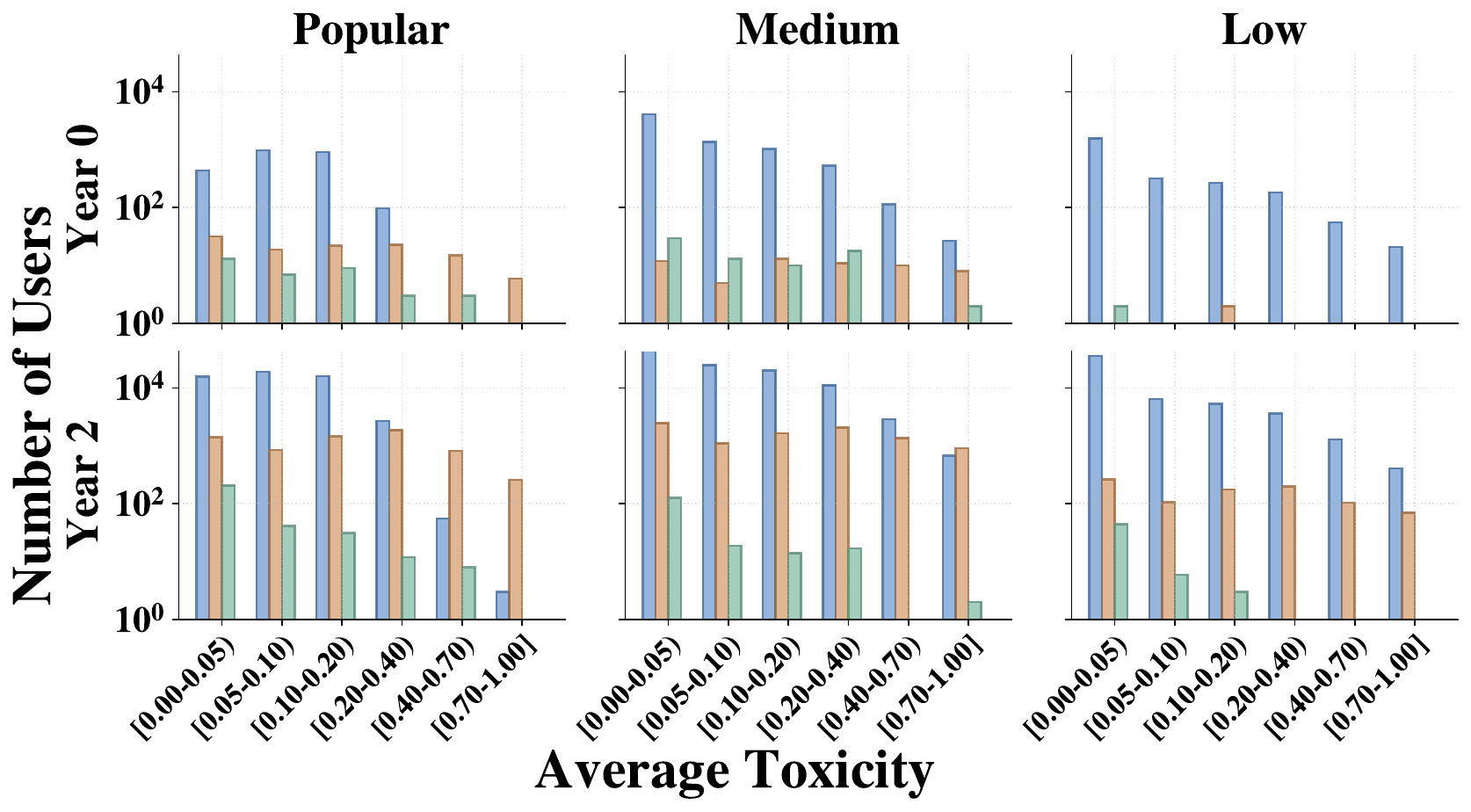}
        \smallskip
        \textbf{(c) Bluesky, Definition I}\label{fig:micro_bs_1}
    \end{minipage}
    \begin{minipage}[t]{0.48\textwidth}
        \centering
        \includegraphics[width=\linewidth]{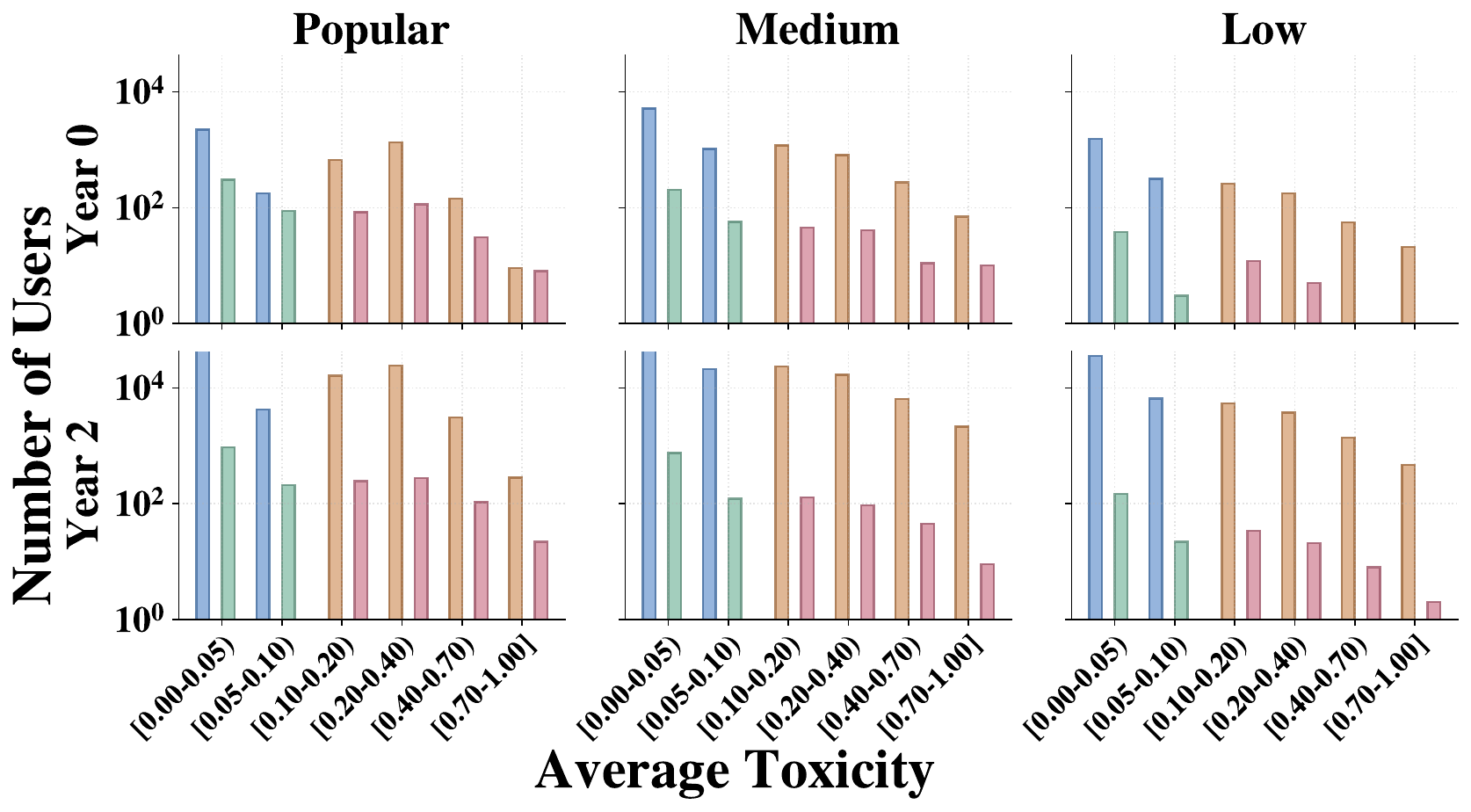}
        \smallskip
        \textbf{(d) Bluesky, Definition II}\label{fig:micro_bs_2}
    \end{minipage}
    \hfill

    \caption{Micro-level toxicity distributions under Definitions I and II on X and Bluesky, using a negative-edge weight of 10. Each panel reports log-scale user counts by average-toxicity bin for intra-/inter-community friends and foes, grouped by user prominence and time snapshot. Overall, friend ties are concentrated in low-toxicity bins, while foe ties are more associated with the high-toxicity tail, indicating localized conflict both within and across communities.}
    \label{fig:micro}
\end{figure*}

\section{Preliminary}\label{sec:data}
We formalize adverse social interactions (ASIs) as time-stamped dyadic events with two complementary channels: structural-based negativity and content-based negativity. Let $G_t=(\mathcal{V}_t,\mathcal{E}_t)$ denote the directed social network at time $t$, where $\mathcal{V}_t$ is the user set and $(u,v)\in\mathcal{E}_t$ indicates that user $u$ follows user $v$. Let $\mathcal{A}^{\mathrm{struct}}=\{(u,v,t)\}$ denote the set of observed structural-based disconnection actions where user $u$ disconnects from user $v$ at time $t$.

\vspace{-0.5ex}
\begin{definition}[Structural-based negativity]
\[
D(u \rightarrow v, t) =
\begin{cases}
1, & \text{if } (u, v, t) \in \mathcal{A}^{\mathrm{struct}}, \\
0, & \text{otherwise}.
\end{cases}
\]
\end{definition}
Let $\mathcal{P}=\{(u,v,p,t)\}$ be the set of directed posts, where user $u$ creates post $p$ directed at user $v$, e.g., through a reply, mention, or quote, at time $t$. Each post $p$ is associated with a toxicity score $\mathrm{Tox}(p)\in[0,1]$ computed by a toxicity classifier.

\vspace{-0.5ex}
\begin{definition}[Content-based negativity]
\[
T(u \rightarrow v)
=
\mathrm{Agg}_{(u,v,p,t)\in\mathcal{P}}
\mathrm{Tox}(p),
\]
\end{definition}
\noindent where $\mathrm{Agg}$ is an aggregation function over directed posts from $u$ to $v$. Here we use the mean toxicity as the default setting.

We collect and curate two large-scale online social network datasets to study ASI across different platform contexts. The first dataset is collected from X and covers 16 weeks of user activity. The second dataset is collected from Bluesky that spans two years. Both datasets contain time-stamped user posts, directed interactions, follow relations, and blocking information, which allow us to align structural-based disconnection events and content-based toxicity into a unified temporal dyadic format. This unified representation enables consistent ASI analysis across platforms and supports the micro-, meso-, and macro-level analyses below. Summary statistics for the datasets are provided in the Appendix.

\section{Micro-level Local Influence of ASI}\label{sec:micro}

We analyze where ASI appears around individual users and how local patterns differ between structural-based and content-based negativity. Specifically, we compare toxicity across four dyadic contexts: intra-community friends, inter-community friends, intra-community foes, and inter-community foes. We conduct this analysis on both platforms under both definitions and further compare patterns across user prominence groups and time snapshots.

\subsection{Setup}

As shown in Figure~\ref{fig:overall}, for each focal user $i$, we combine community membership $c(i)$ with dyadic friend/foe status. Friends are social ties, while foes are ASI ties under the chosen definition. For each of the four groups, we compute the average toxicity of interactions involving $i$ and aggregate the distributions by prominence tier and time snapshot. Communities are inferred from signed network structure with a fixed negative-edge weight of 10, while toxicity is measured only after the partition is fixed. Thus, the intra- versus inter-community toxicity comparison is not built into the clustering.

\subsection{Result Analysis and Insights}

We summarize the main micro-level findings from Figure~\ref{fig:micro}:

\vspace{0.5ex}
\noindent\textbf{Platform context affects where ASI is locally situated.}
Both X and Bluesky show a clear separation between friend and foe ties: friend ties are concentrated in the lowest toxicity bins, while foe ties extend further into medium- and high-toxicity bins. The more informative difference lies in how these ties are positioned relative to community boundaries. On X, especially under Definition II in Figure~\ref{fig:micro}(b), inter-community categories remain visible across prominence groups, suggesting that local conflict on X contains a stronger boundary-spanning component. On Bluesky, especially under Definition I in Figure~\ref{fig:micro}(c), the distributions are more dominated by intra-community friends and intra-community foes, indicating that structurally negative relations are more embedded within detected community contexts. Thus, the two platforms share the same basic friend--foe toxicity separation, but differ in the local community placement of ASI.

\vspace{0.5ex}
\noindent\textbf{The two ASI definitions expose different dimensions of negative interaction.}
Definition I uses structural disconnection to identify foe ties. In Figures~\ref{fig:micro}(a) and~\ref{fig:micro}(c), these structurally negative ties are more represented in higher-toxicity bins than friend ties. This shows that blocking and unfollowing identify dyadic contexts where hostile communication is more likely to appear, even though the definition itself does not use toxicity. Definition II uses toxic directed communication to identify foe ties. In Figures~\ref{fig:micro}(b) and~\ref{fig:micro}(d), the key information is how toxic ties are distributed across intra- and inter-community positions. These panels show that content-based ASI appears both inside communities and across community boundaries. Together, the two definitions provide complementary signals: structural-based ASI captures relational breakdown, while content-based ASI captures hostile communication and its local community placement.

\vspace{0.5ex}
\noindent\textbf{The temporal snapshots show persistent friend--foe separation with stronger intra-community concentration in later periods.}
Friend ties remain concentrated in the lowest toxicity bins across time snapshots on both platforms, and foe ties continue to occupy a wider toxicity range. This repeated pattern suggests that the micro-level organization of ASI is not limited to a single snapshot. The later snapshots also show strong visibility of intra-community foe ties in several settings, especially under Definition II. Since the figure reports counts, these changes should be interpreted descriptively rather than as normalized growth rates. Still, the temporal comparison indicates that adverse ties remain locally concentrated over time and continue to appear within existing community contexts.

\vspace{0.5ex}
\noindent\textbf{High-toxicity interactions are concentrated around specific adverse dyads.}
Across platforms, definitions, prominence groups, and time snapshots, friend ties are concentrated in lower-toxicity bins, while foe ties account for more of the medium- and high-toxicity range. This pattern is especially meaningful under Definition I, where toxicity is not part of the foe-tie construction. The result suggests that micro-level negativity is highly localized: most observed social ties involve low average toxicity, while a smaller set of adverse dyads carries a disproportionate share of severe toxic interaction. This supports a dyadic interpretation of ASI, where conflict is organized around particular user pairs rather than evenly distributed across a user's local neighborhood.

\vspace{0.5ex}
\noindent\textbf{ASI reflects both internal community friction and boundary conflict.}
Intra-community foe ties are consistently visible across panels, showing that adverse interactions often occur among users assigned to the same detected community. This points to internal community friction: shared community membership does not imply uniformly benign interaction. Inter-community foe ties are also present, especially under content-based ASI, and capture hostility that crosses community boundaries. The figure therefore suggests two local forms of ASI. Some adverse interactions may weaken cohesion within communities, while others may sharpen conflict between communities.

\vspace{0.5ex}
\noindent\textbf{User prominence changes the scale of observation, not the core pattern.}
Popular users generally show broader distributions and larger counts, which is consistent with their larger interaction volume and more heterogeneous audiences. Medium- and low-prominence users show the same qualitative structure: friend ties remain concentrated in low-toxicity bins, while foe ties are more visible in the higher-toxicity range. This indicates that the micro-level organization of ASI is not driven only by highly visible users. Prominence increases exposure and heterogeneity, but the friend--foe toxicity pattern appears across all prominence groups.

\section{Meso-Level Causal Influence of ASI}

Our central question is whether ASI diffuses through follower ties in a causal sense: after a source user $U_s$ takes a ASI toward a target $U_t$, does an exposed follower $U_e$ become more likely to take the same act toward $U_t$? 
This meso-level mechanism links micro acts to macro structural change and complements our micro and macro analyses.
We identify this effect via \emph{balanced risk-set matching}, pairing exposed triads with comparable unexposed configurations within the same time context~\cite{rosenbaum1983central, rosenbaum2010design, russo2024stranger}.

\subsection{Setup}
\label{sec:setup-notation}

For Bluesky dataset, let $\mathcal{P}$ denote public interactions (reply, mention, quote) with timestamps, and let $\mathcal{B}$ denote block events with timestamps; our dataset includes per-post text/timing and time-stamped follows/blocks. Throughout this section we use triads $(U_{\mathrm{e}}, U_{\mathrm{s}}, U_{\mathrm{t}})$, where $U_{\mathrm{s}}$ is the \emph{source} who performs an initial ASI toward a \emph{target} $U_{\mathrm{t}}$,
and $U_{\mathrm{e}}$ is an \emph{exposed follower} (a user who follows $U_{\mathrm{s}}$ prior to exposure and may subsequently act toward $U_{\mathrm{t}}$). 
We write $T^\star$ for the exposure time associated with $U_{\mathrm{s}}$'s act.

\begin{figure}[t]
  \centering
  \includegraphics[width=0.95\columnwidth]{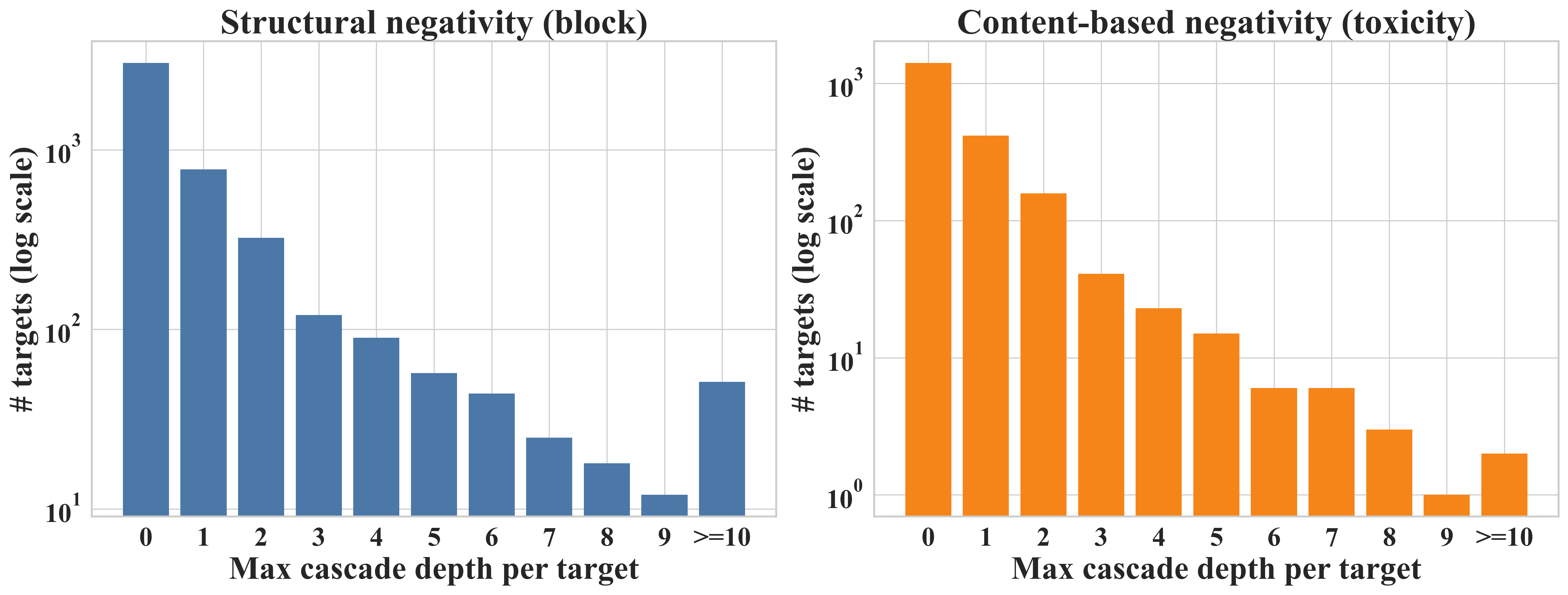}
  \vspace{-1.75ex}
  \caption{Depth distribution of cascading trees (log-scale $y$) on Bluesky. For each target $U_t$, we plot the histogram of the maximum cascade depth within a $W_{\mathrm{cas}}{=}7$ day window, for (left) block events and (right) toxic interactions.}
  \label{fig:cascade-depth-baseline}
  \vskip -3ex
\end{figure}

\paragraph{Descriptive cascade prevalence (cascading trees).}
Before turning to causal identification, we perform an initial descriptive check for diffusion patterns by constructing cascading trees centered on each target $U_t$.
Let $W_{\mathrm{cas}}$ denote the cascade linking window (default $7$ days).
For each target $U_t$, we construct a directed cascade among users who act against $U_t$ by linking an earlier actor $U_s$ to a later actor $U_e$ whenever $U_e$ was exposed to $U_s$ through a pre-existing follow tie (i.e., $U_e$ followed $U_s$ prior to $U_s$'s act) and performs the same act shortly after $U_s$ (within $W_{\mathrm{cas}}$).
To obtain a forest structure and a well-defined depth, we assign each $U_e$ to at most one parent $U_s$, chosen as the most recent eligible predecessor within $W_{\mathrm{cas}}$.
We summarize these trees using depth statistics.

Our findings in Figure~\ref{fig:cascade-depth-baseline} show a highly skewed, long-tailed depth distribution: while many targets have $\texttt{max\_depth}=0$, a non-trivial set exhibits multi-hop cascades (e.g., $\texttt{max\_depth}\ge 2$), motivating a closer, causal investigation of follower-based diffusion.

A \textbf{treated triad} $(U_{\mathrm{e}},U_{\mathrm{s}},U_{\mathrm{t}})$ satisfies:
(i) $U_{\mathrm{e}}$ follows $U_{\mathrm{s}}$ at some $t_0<T^\star$;
(ii) $U_{\mathrm{s}}$ publicly interacts with $U_{\mathrm{t}}$ at $t_{\mathrm{int}}\le T^\star$; and
(iii) $U_{\mathrm{s}}$ blocks $U_{\mathrm{t}}$ by $t_{\mathrm{blk}}\in (t_{\mathrm{int}}, t_{\mathrm{int}}+\Delta]$,
with exposure time $T^\star\coloneqq t_{\mathrm{blk}}$ and default $\Delta=3$ days.
Exposure assumes $U_{\mathrm{e}}$ could observe (or be informed by) $U_{\mathrm{s}}$’s public interaction prior to the block.

\paragraph{Outcome.}
Fix an outcome window $W_{\mathrm{out}}>0$. Define
\[
Y_T(U_{\mathrm{e}},U_{\mathrm{s}},U_{\mathrm{t}}) \coloneqq \mathds{1}\!\Big\{ U_{\mathrm{e}} \text{ blocks } U_{\mathrm{t}} \text{ in } (T^\star,\, T^\star + W_{\mathrm{out}}] \Big\}.
\]

\paragraph{Pre-treatment text embeddings for matching.}
For any user $u$, compute a pre-treatment embedding $z_u\in\mathbb{R}^d$~\cite{reimers2019sentence} by encoding all posts in $[T^\star - W_{\mathrm{pre}},\, T^\star)$ and mean-pooling, then L2-normalize and compare with cosine similarity~\cite{veitch2020adapting}. We require a minimal pre-treatment footprint (e.g., $\ge 50$ posts when available) and cap long histories per dataset constraints.

\paragraph{Triadic replacement principle (what changes vs.\ what is fixed).}
Our causal question is meso-level and triadic. We isolate the channel by replacing one user while holding the other two fixed~\cite{rosenbaum1983central, stuart2010matching, li2001balanced}:
\[
\begin{aligned}
\mathsf{Rep}_t(U_e,U_s,U_t;\mathcal{C}_t) &\coloneqq (U_e, U_s, U_t'),\\
\mathsf{Rep}_e(U_e,U_s,U_t;\mathcal{C}_e) &\coloneqq (U_e', U_s, U_t),
\end{aligned}
\]
where $\mathcal{C}_t$ and $\mathcal{C}_e$ are admissibility constraints (pre-exposure similarity, exposure availability, and no disqualifying pre-ties). Intuitively, $\Rep_t$ yields a “changed target” triad (same $U_e,U_s$), while $\Rep_e$ yields a “changed exposed” triad (same $U_s,U_t$).

\begin{figure}[t]
  \centering
  \includegraphics[width=0.95\linewidth]{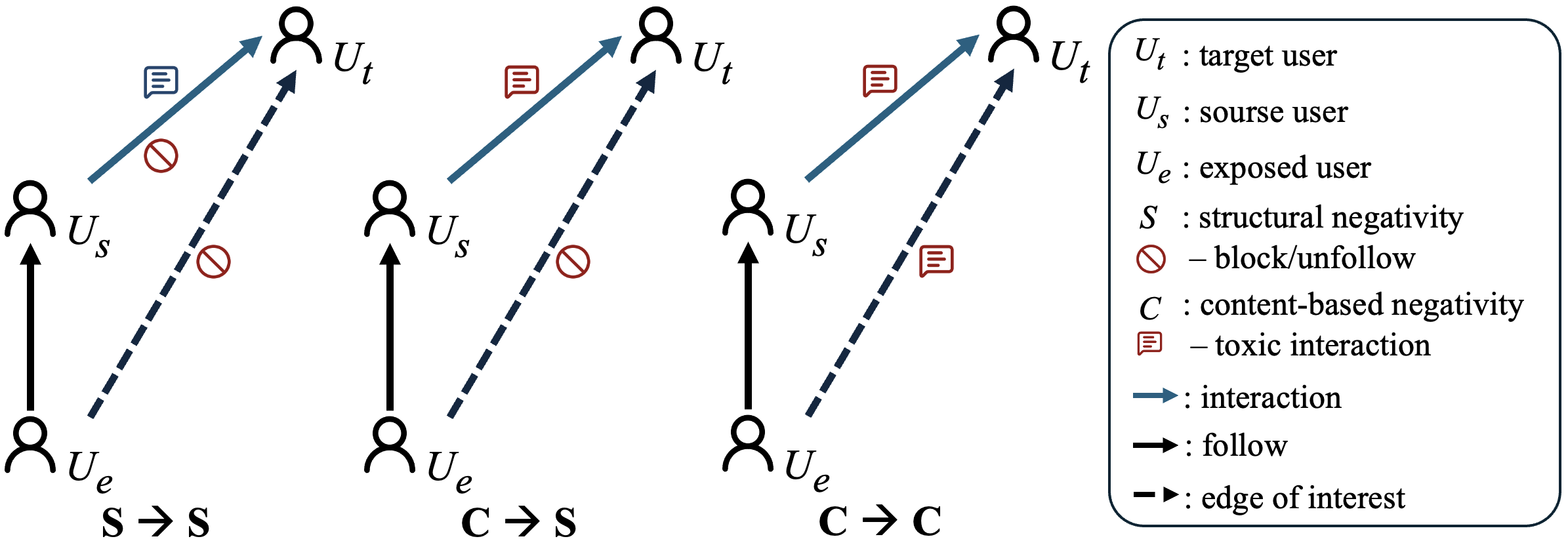}
  \caption{\textbf{Structure Toxicity Coupling.}
  Illustration of how we couple structural relations (follow edges) with
  toxicity signals when defining exposure and outcome links.}
  \label{fig:struct-tox}
\end{figure}

\subsection{Potential Outcomes, Estimands, and Assumptions}
For a treated triad $(U_e,U_s,U_t)$ with exposure time $T^\star$, define the replacement-based controls
$C^{(t)}=\mathsf{Rep}_t(U_e,U_s,U_t;\mathcal{C}_t)$ and
$C^{(e)}=\mathsf{Rep}_e(U_e,U_s,U_t;\mathcal{C}_e)$.
Let $Y_C^{(t)}$ and $Y_C^{(e)}$ be the corresponding control outcomes, measured in the same window $(T^\star,\,T^\star+W_{\mathrm{out}}]$ as in \S\ref{sec:setup-notation}. 

\noindent We estimate the paired risk differences
\[
\tau^{(t)}=\mathbb{E}\!\big[\,Y_T - Y_C^{(t)}\,\big],\qquad
\tau^{(e)}=\mathbb{E}\!\big[\,Y_T - Y_C^{(e)}\,\big],
\]
reported separately and, where appropriate, pooled via fixed-effects meta-analysis across platforms/settings.

\emph{Identification.} Assume consistency; no cross-pair interference~\cite{hudgens2008toward} within $W_{\mathrm{out}}$; and conditional ignorability with overlap~\cite{rubin2005causal, rosenbaum2010design}:
\[
(Y_T, Y_C^{(t)}, Y_C^{(e)})\;\perp\!\!\!\perp\;\mathrm{Treatment}\;\big|\;\mathrm{pair}(U_e,U_s,U_t;\mathcal{C}_t,\mathcal{C}_e).
\]
The constraints $\mathcal{C}_t,\mathcal{C}_e$ (embedding similarity, proximity windows, tie-strength strata) are chosen to block backdoor paths between replacement choice and outcomes.

\subsection{Design Overview: Replacement \texorpdfstring{$\times$}{\texttimes} Modality}
\label{sec:design-overview}
As visualized in Figure~\ref{fig:struct-tox}, we cross two replacement designs—\emph{Match-$U_t$} (changed target) and \emph{Match-$U_e$} (changed exposed)—with three exposure$\to$outcome couplings:
\begin{itemize}[leftmargin=1.2em,itemsep=0.2em,topsep=0.1em]
  \item \textbf{S$\to$S} (structure$\to$structure): $U_s$ publicly interacts with $U_t$ and then disconnects; outcome: whether $U_e$ disconnects $U_t$.
  \item \textbf{C$\to$S} (content$\to$structure): a $U_s\!\to\!U_t$ post exceeds a toxicity threshold; outcome: whether $U_e$ disconnects $U_t$.
  \item \textbf{C$\to$C} (content$\to$content): toxic exposure $U_s\!\to\!U_t$; outcome: whether $U_e$ produces toxic content toward $U_t$ (global toxicity score thresholds).
\end{itemize}

\vspace{-0.75ex}
\paragraph{Why two designs?}
Match-$U_t$ tests whether $U_s$’s act is target-specific by varying the target while holding $(U_e,U_s)$ fixed; Match-$U_e$ tests whether exposure via following $U_s$ matters by varying the exposed user while holding $(U_s,U_t)$ fixed.

\vspace{-0.75ex}
\paragraph{Operational choices shared across cells.}
All cells use balanced risk-set matching at $T^\star$ with (i) pre-exposure text-embedding similarity, (ii) proximity rules around $U_s$, and (iii) exclusion of disqualifying pre-ties. We fix $(\Delta, W_{\mathrm{pre}}, W_{\mathrm{out}}, w)$ across cells for comparability, and analyze outcomes with McNemar on discordant pairs with exact CIs. See \S\ref{sec:treat-control} for operational details.

\begin{figure}[t]
  \centering
  \includegraphics[width=0.9\columnwidth]{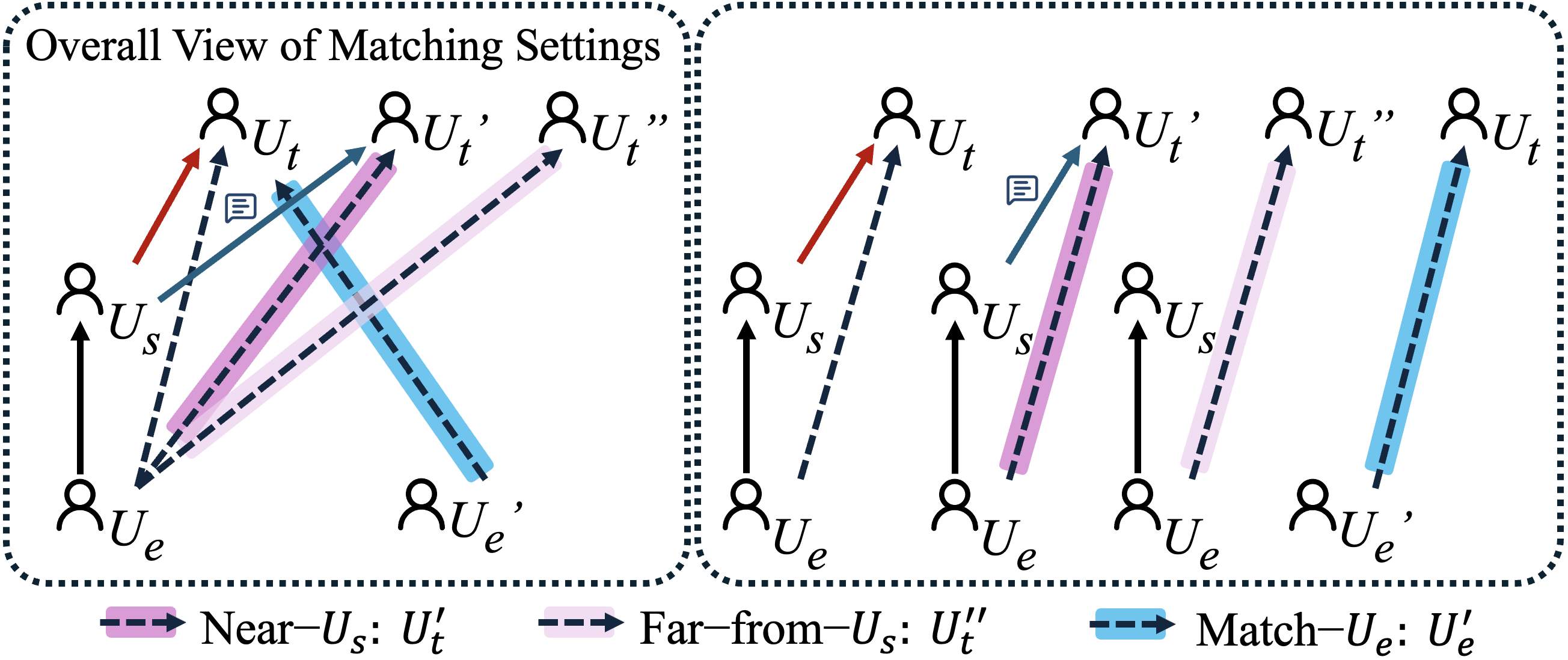}
  \vspace{-2ex}
  \caption{\textbf{Matching Setting.}
  Overall view of our matching designs. Left: joint view of the main
  triad $(U_{\mathrm{e}}, U_{\mathrm{s}}, U_{\mathrm{t}})$ and the
  replacement targets $U_{\mathrm{t}}', U_{\mathrm{t}}''$ and exposed
  users $U_{\mathrm{e}}'$. Right: separated panels highlighting each
  matching configuration used in our experiments.}
  \label{fig:matching-settings}
  \vspace{-3ex}
\end{figure}

\vspace{-0.75ex}
\subsection{Treatment/Control: Operationalizing Replacement}
\label{sec:treat-control}
We specify (i) admissibility constraints $(\mathcal{C}_t,\mathcal{C}_e)$, (ii) shared windows $(\Delta, W_{\mathrm{pre}}, W_{\mathrm{out}}, w)$, and (iii) pair-construction rules for both replacement designs across S$\to$S, C$\to$S, and C$\to$C.

\vspace{-0.75ex}
\paragraph{Negative-behavior settings.}
\textsc{Struct}: disconnection action indicator $D(u\!\to\!v,t)$; \textsc{Content}: directed public interaction with toxicity $T(u\!\to\!v)\!\ge\!\tau$ (or user-specific $\tau_u$). Results 
reported platform-wise.

\vspace{-0.75ex}
\paragraph{Exposure time and outcomes.}
\begin{itemize}[leftmargin=1.2em,itemsep=0.2em]
\item \textsc{Struct}: $T^\star=t_{\mathrm{blk}}$ if $U_s$ interacts with $U_t$ at $t_{\mathrm{int}}\le T^\star$ and disconnects within $\Delta$; outcome: if $U_e$ blocks $U_t$ in $(T^\star, T^\star+W_{\mathrm{out}}]$.
\item \textsc{Content}: exposure when a $U_s\!\to\!U_t$ post crosses $\tau$ (or $\tau_u$); outcomes: (a) $U_e$ blocks $U_t$ (C$\to$S) or (b) $U_e$ produces toxic content toward $U_t$ (C$\to$C) within the same window.
\end{itemize}

\vspace{-0.75ex}
\paragraph{Pair constructions (fix two of the triad). We show the overview of our matching setting in Figure~\ref{fig:matching-settings}.}
\begin{enumerate}[leftmargin=1.2em,itemsep=0.2em]
\item \textbf{Match-$U_{\mathrm{t}}$} (hold $U_{\mathrm{e}},U_{\mathrm{s}}$). 
Let $\operatorname{Int}_B(u;W_{\mathrm{int}})$ be the \# public interactions between $U_s$ and $u$ in $[T^\star\!-\!W_{\mathrm{int}},\,T^\star)$. Select $U_t'$ (or $U_t''$) maximizing $\cos(z_C,z_u)$ subject to:
\begin{enumerate}[label=(\alph*), leftmargin=1.5em,itemsep=0.2em]
\item \emph{Near-$U_s$ ($U_t'$):} $u$ interacted with $U_s$ in-window and $\operatorname{Int}_B(u;W_{\mathrm{int}}) > \operatorname{Int}_B(U_t;W_{\mathrm{int}})$; $U_s$ has not blocked $u$ by $T^\star$.
\item \emph{Far-from-$U_s$ ($U_t''$):} $\operatorname{Int}_B(u;W_{\mathrm{int}})=0$ (and not blocked by $T^\star$).
\end{enumerate}

\item \textbf{Match-$U_{\mathrm{e}}$} (hold $U_{\mathrm{s}},U_{\mathrm{t}}$).
Pick $U_{\mathrm{e}}'$ maximizing $\cos(z_A,z_{A'})$ with no $U_{\mathrm{e}}'\!\to\!U_s$ follow in $[T^\star\!-\!w,\,T^\star\!+\!w]$ and no $U_{\mathrm{e}}'\!\leftrightarrow\!U_t$ interaction in $[T^\star\!-\!W_{\mathrm{pre}},\,T^\star)$.
\end{enumerate}

\paragraph{Strata and case-study variants (tie strength).} We also 
examine how pre-existing relationships between
$U_{\mathrm{e}}$ and $U_{\mathrm{s}}$ modify the cascade effect in a focused
case-study setting. For a given triad and exposure time $T^\star$, let
$N^{30}(U_{\mathrm{e}},U_{\mathrm{s}})$ denote the number of directed public
interactions from $U_{\mathrm{e}}$ to $U_{\mathrm{s}}$ in
$[T^\star - 30\text{d},\,T^\star]$. This provides a simple tie-strength measure for the exposed follower
and source.

In a representative design cell, we first construct matched
treated \& control pairs as in \S\ref{sec:treat-control}, then label each
pair by $N^{30}(U_{\mathrm{e}},U_{\mathrm{s}})$ and split the pairs into
\textsc{High} and \textsc{Low} strata via an upper-quantile cut
(e.g., above vs.\ below the $80$th percentile of $N^{30}$ among
eligible followers of the same source).

\paragraph{Multi-source exposure case study (number of sources).}
We also ask whether cascades strengthen when the exposed follower
is linked to \emph{multiple} sources who have already acted negatively
toward the same target. For a target $U_{\mathrm{t}}$, let
$\mathcal{S}(U_{\mathrm{t}})$ be the set of users who block
$U_{\mathrm{t}}$ with associated block times $\{(t_b,U_{\mathrm{s}})\}$,
sorted by $t_b$. For any follower $U_{\mathrm{e}}$ and integer
$m \ge 1$, we define a multi-source exposure time $T^\star_m$ as the
time of the $m$-th block of $U_{\mathrm{t}}$ by a user in
$\mathcal{S}(U_{\mathrm{t}})$ that $U_{\mathrm{e}}$ follows, i.e., the
earliest time at which $U_{\mathrm{e}}$ follows $m$ distinct members of
$\mathcal{S}(U_{\mathrm{t}})$ who have blocked $U_{\mathrm{t}}$ by
$T^\star_m$. We drop $(U_{\mathrm{e}},U_{\mathrm{t}})$ if
$U_{\mathrm{e}}$ blocked $U_{\mathrm{t}}$ before $T^\star_m$.
In this case study we focus on heavily targeted users, restricting to
the most frequently blocked $U_{\mathrm{t}}$ (top $K$ by number of
blockers).

For a given design cell and value of $m$, each $(U_{\mathrm{e}},U_{\mathrm{t}})$
with valid $T^\star_m$ is treated at level $X \ge m$ and matched to
controls $(U_{\mathrm{e}}',U_{\mathrm{t}})$ who (i) follow none of the
blockers of $U_{\mathrm{t}}$ and (ii) have not blocked $U_{\mathrm{t}}$
by $T^\star_m$. Matching again uses pre-exposure text embeddings and a
risk-set constraint that restricts controls to users who have not yet
blocked $U_{\mathrm{t}}$ at $T^\star_m$, and we estimate the paired risk
difference in $U_{\mathrm{e}}$ vs.\ $U_{\mathrm{e}}'$ blocking
$U_{\mathrm{t}}$ within the outcome window.

\subsection{Estimator and Test}
\label{sec:estimator}
With paired Bernoulli outcomes $(Y_T,Y_C)\in\{0,1\}^2$ across $N$ pairs, let $a,b,c,d$ be the counts of $(1,1)$, $(1,0)$, $(0,1)$, $(0,0)$, respectively. The paired risk difference is
\vspace{-0.5ex}
\[
\hat{\tau} \;=\; \frac{1}{N}\sum (Y_T - Y_C) \;=\; \frac{b-c}{N}.
\]

\vspace{-0.5ex}
\noindent We test $H_0:\Pr(Y_T=1)=\Pr(Y_C=1)$ using McNemar~\cite{mcnemar1947note} on discordant pairs: a continuity-corrected statistic
\[
\chi^2 \;=\; \frac{(\lvert b-c\rvert-1)^2}{b+c}
\]
and the exact binomial alternative with $b \sim \operatorname{Binom}(b+c,\, 0.5)$. We also report the paired odds ratio $\widehat{\theta}=b/c$ with exact (mid-$p$) confidence intervals. 
Pair construction Algorithm~\ref{alg:pair-construction} is in Appendix.

\begin{table}[t]
\centering
\scriptsize
\caption{Combined summary across replacement settings and exposure$\to$outcome couplings. X rows use $W{=}1$ only; Count shows approximate sample-size bin. Values are Treated rate (per 1k), Times ($\times$=Treated/Control), and significance from McNemar exact $p$.
} 
\vspace{-1.75ex}
\label{tab:meso-summary}
\resizebox{\columnwidth}{!}{
\begin{tabular}{lllcccc}
\toprule
\makecell[l]{Replacement\\Setting} &
\makecell[l]{Coupling\\Method} &
\makecell[l]{Social\\Platform} &
\makecell[c]{Pair\\Count} &
\makecell[c]{Treated rate\\(per 1k)} &
\makecell[c]{$\times$\\(T/C)} &
\makecell[c]{Sig} \\
\midrule
\multirow{5}{*}{%
  \makecell[l]{%
    \textit{Match-$U_{\mathrm{t}}'$}\\[-0.3ex]
    \scriptsize\textit{(changed target,}\\[-0.3ex]
    \scriptsize\textit{near-$U_s$)}%
  }%
}
 & S$\to$S & Bluesky & $>{}100$k & 1.118 & 24.3$\times$ & *** \\
 & C$\to$S & Bluesky & $>{}1$M    & 0.175 & 3.6$\times$  & *** \\
 & C$\to$C & Bluesky & $>{}1$M    & 2.803 & 0.8$\times$  & *** \\
 & C$\to$S & X  & $>{}10$k   & 0.156 & 0.5$\times$  & ns \\
 & C$\to$C & X  & $>{}10$k   & 3.769 & 0.7$\times$  & ** \\
\addlinespace
\multirow{5}{*}{%
  \makecell[l]{%
    \textit{Match-$U_{\mathrm{t}}''$}\\[-0.3ex]
    \scriptsize\textit{(changed target,}\\[-0.3ex]
    \scriptsize\textit{far from-$U_s$)}%
  }%
}
 & S$\to$S & Bluesky & $>{}100$k & 1.107 & 29.1$\times$ & *** \\
 & C$\to$S & Bluesky & $>{}1$M    & 0.165 & 1.4$\times$  & *** \\
 & C$\to$C & Bluesky & $>{}1$M    & 3.059 & 1.3$\times$  & *** \\
 & C$\to$S & X  & $>{}10$k   & 0.587 & 4.5$\times$  & ** \\
 & C$\to$C & X  & $>{}10$k   & 5.698 & 1.1$\times$  & ns \\
\addlinespace
\multirow{5}{*}{%
  \makecell[l]{%
    \textit{Match-$U_{\mathrm{e}}'$}\\[-0.3ex]
    \scriptsize\textit{(changed}\\[-0.3ex]
    \scriptsize\textit{exposed follower)}%
  }%
}
 & S$\to$S & Bluesky & $>{}100$k & 1.032 & 3.5$\times$ & *** \\
 & C$\to$S & Bluesky & $>{}1$M    & 0.163 & 1.6$\times$ & *** \\
 & C$\to$C & Bluesky & $>{}1$M    & 2.989 & 2.0$\times$ & *** \\
 & C$\to$S & X  & $>{}10$k   & 0.499 & 1.0$\times$ & ns \\
 & C$\to$C & X  & $>{}10$k   & 3.794 & 2.7$\times$ & *** \\
\bottomrule
\end{tabular}}
\vspace{2pt}
\begin{minipage}{\columnwidth}
\scriptsize
Notes: Sig legend (McNemar exact $p$): ns $\ge .05$, * $< .05$, ** $< .01$, *** $< .001$.\\
Treated rate is reported per 1,000 (i.e., raw rate $\times 10^3$).\\
X rows report $W{=}1$ only (W2/W3 omitted by design). Bluesky has no windows. X S$\to$S unavailable due to sparse unfollow events.
\end{minipage}
\vskip -4ex
\end{table}

\subsection{Meso-level Result Analysis}
\label{sec:meso-results}

Table~\ref{tab:meso-summary} summarizes paired causal estimates across
two replacement strategies (Match-$U_t$ vs.\ Match-$U_e$) and three
exposure$\to$outcome couplings (S$\to$S, C$\to$S, C$\to$C).
Because ASI outcomes are rare, we report both multiplicative
effects (T/C) and emphasize absolute paired risk differences implied by treated--control rates. The detailed results on both datasets are shown in Table~\ref{tab:bluesky_original} and~\ref{tab:X_original} in Appendix.

\vspace{0.5ex}
\textbf{Strong, target-specific structural diffusion on Bluesky (S$\to$S).}
  Across both target replacements (near-$U_s$ and far-from-$U_s$), exposure to $U_s$'s interaction$\to$block toward $U_t$ \emph{causally} increases the probability that an exposed follower $U_e$ also blocks $U_t$:
  treated $\approx 1.1$ per 1k with very large T/C ratios (24--29$\times$), and Match-$U_e$ remains significantly positive (3.5$\times$).
  This consistency across both replacement designs supports a robust meso-level ``follow-the-source'' mechanism for structural disconnection.

\vspace{0.5ex}
\textbf{Content-driven diffusion is weaker and more platform-sensitive (C$\to$S \& C$\to$C).}
    On Bluesky, C$\to$S is consistently positive but small (0.16--0.18 per 1k; 1.4--3.6$\times$), while C$\to$C is positive mainly under Match-$U_e$ (2.0$\times$) and more mixed under Match-$U_t$.
    On X, content effects are more sensitive to counterfactual construction: C$\to$S can be null for near-$U_s$ but positive for far-from-$U_s$, and C$\to$C is strongest under Match-$U_e$ (2.7$\times$).

\begin{figure}[t]
  \centering
  \includegraphics[width=\columnwidth]{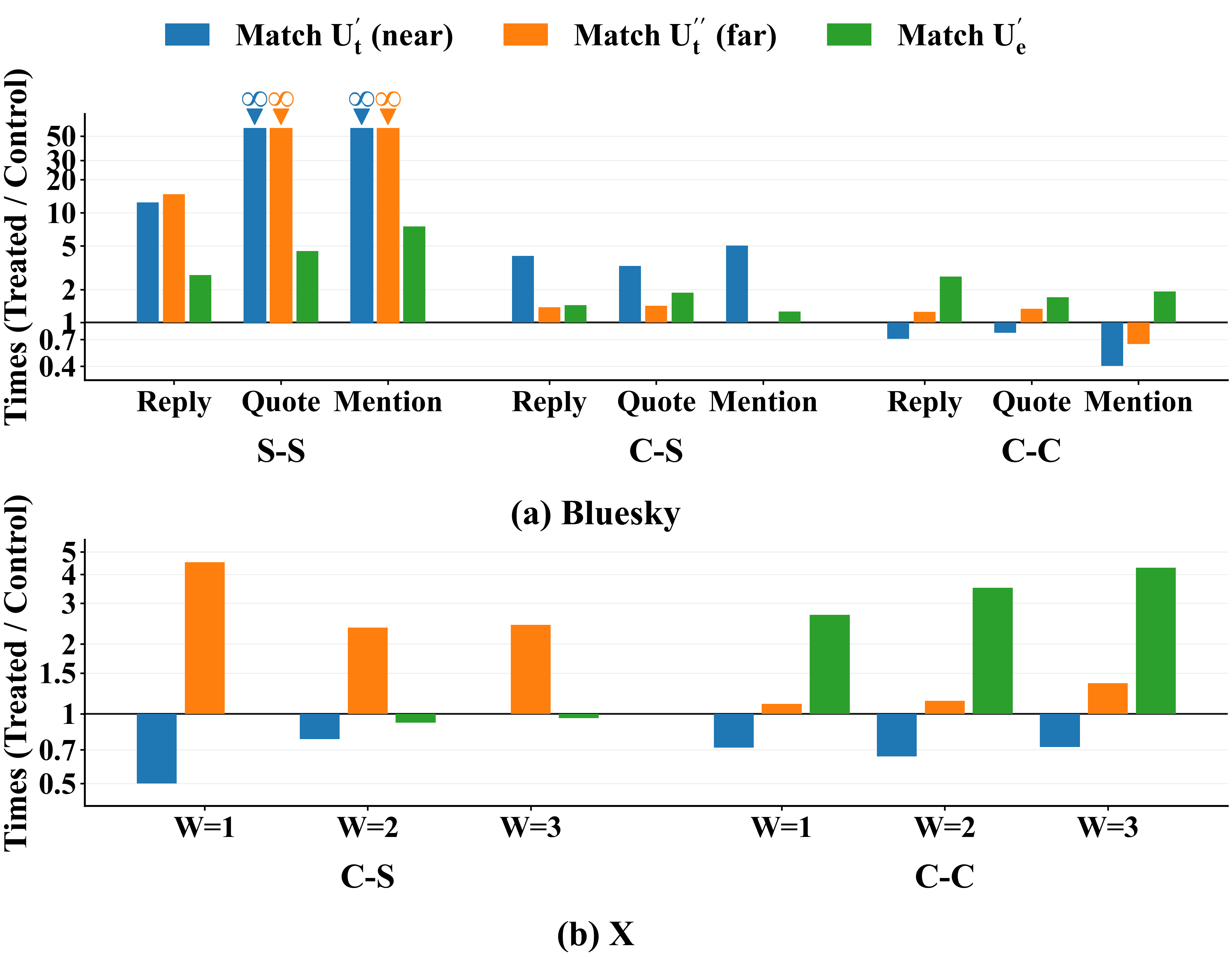}
  \vspace{-4ex}
  \caption{Meso-level causal effects (Times = Treated/Control) across matching settings and platforms. We group the effects of the interaction type in (a) on Bluesky and (b) is about causal delay on X.}
  \label{fig:group_results}
    \vspace{-1.5ex}
\end{figure}

\vspace{1ex}
\textbf{Heterogeneity by interaction type and causal delay.}
    Figure~\ref{fig:group_results} breaks effects down by interaction type and by the outcome window.
On Bluesky, S$\to$S remains consistently positive across interaction types, with a generally larger effect for mentions than quotes, and for quotes than replies; in contrast, C$\to$S and C$\to$C are smaller and more interaction-dependent.
On X, effects extend beyond $W{=}1$, with the clearest persistence under Match-$U_e$ for C$\to$C, consistent with delayed follower responses to toxic exposure.

\begin{table}[t]
  \centering
  \small
  \caption{Tie-strength case study for Match-$U_{\mathrm{e}}$ with S$\to$S coupling on Bluesky.
We stratify matched pairs by the number of prior exposed--source interactions in the 30 days before exposure
($N^{30}(U_{\mathrm{e}},U_{\mathrm{s}})$): \textsc{High} (top 20\%) vs.\ \textsc{Low}.}
\vspace{-1ex}
  \begin{tabular}{lccccc}
    \toprule
    Stratum & $n$ & Treated rate & Control rate & $\times$ (T/C) & $\Delta$ \\
    \midrule
    High tie & 9{,}516   & 0.003468 & 0.000420 & $8.26\times$ & 0.003047 \\
    Low tie  & 147{,}427 & 0.000875 & 0.000292 & $3.00\times$ & 0.000583 \\
    \midrule
    \multicolumn{6}{l}{\footnotesize
    Test for effect modification:
    $\Delta_H - \Delta_L = 0.002464$, $z=3.823$, $p=1.32\times10^{-4}$,} \\
    \multicolumn{6}{l}{\footnotesize
    \hspace{1.7em}95\% CI $[0.001201,\,0.003728]$.}\\
    \bottomrule
  \end{tabular}
  \label{tab:tie-strength}
  \vskip -1ex
\end{table}

\begin{figure}[t]
    \centering
    \includegraphics[width=0.9\columnwidth]{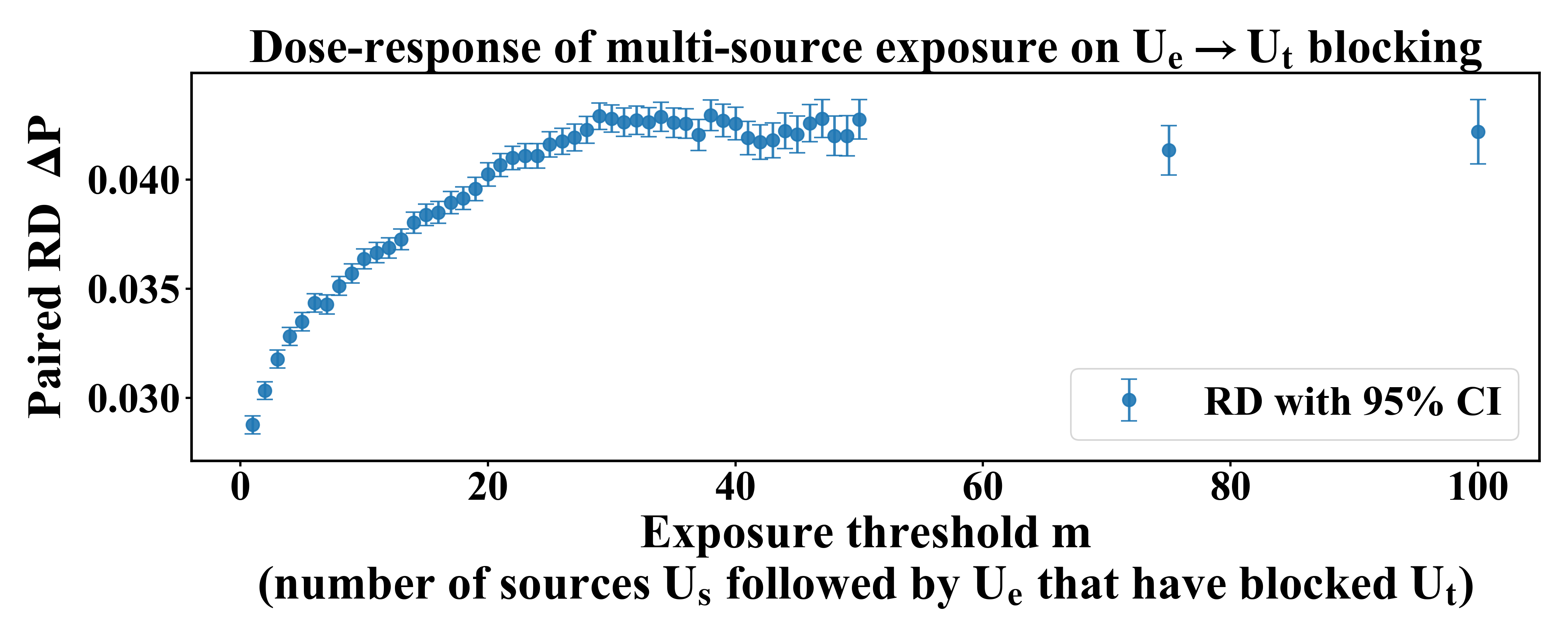}
    \vspace{-2.5ex}
    \caption{Dose--response of multi-source exposure on $U_e{\rightarrow}U_t$ blocking.
    For each exposure threshold $m$, we consider heavily targeted users $U_t$ and exposed followers $U_e$ who follow at least $m$ distinct sources $U_s$ that have blocked $U_t$ by $T^\star_m$.
    Treated pairs $(U_e,U_t)$ with exposure level $X \ge m$ are matched to controls $(U_e',U_t)$ who follow none of $U_t$'s blockers and have not blocked $U_t$ by $T^\star_m$.
    The curve reports paired risk difference $\Delta P(U_e \text{ blocks } U_t)$ between treated and control. }
    \label{fig:multi_source_dose}
    \vspace{-1.5ex}
\end{figure}

\vspace{1ex}
\textbf{Effect modification by tie strength.}
    We test whether diffusion is stronger along closer follower--source ties by stratifying matched pairs.
Table~\ref{tab:tie-strength} shows clear effect modification: the \textsc{High} stratum (top 20\%) has a larger treated/control ratio ($8.26\times$) than the \textsc{Low} stratum ($3.00\times$).
The difference in risk differences is significant, indicating amplification along stronger ties.

\vspace{1ex}
\textbf{Dose--response under multi-source exposure.}
    We test a reinforcement mechanism by varying the number of distinct followed sources who have already acted against the same target.
Figure~\ref{fig:multi_source_dose} shows that the paired risk difference increases with the exposure threshold $m$, which is consistent with cumulative social exposure amplifying follower adoption of the same act.

\section{Macro-level Community Impact of ASI}\label{sec:macro}

Macro-level analysis is designed to understand the evolutionary impact of ASI on the cohesive subgroups' disruptions in social networks. 

\vspace{-1.5ex}
\subsection{Setup}
\begin{figure}[t]
    \centering
    \includegraphics[width=1\columnwidth]{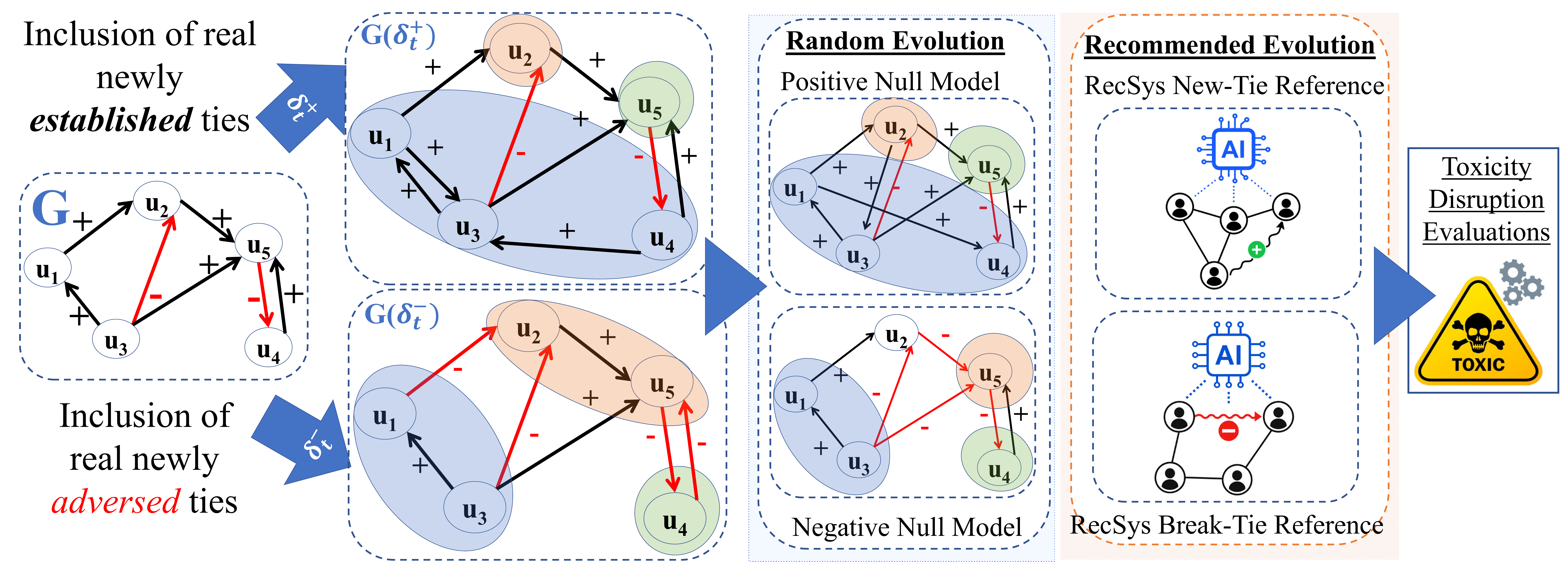}
    \vspace{-3.5ex}
    \caption{Macro-level setup for comparing observed ASI evolution with Random and RecSys references.}
\label{fig:macro-setup}
\vskip -2ex
\end{figure}

We frame macro-level ASI impact as a proxy-based counterfactual analysis of cohesive subgroup disruption. As shown in Figure~\ref{fig:macro-setup}, for each network-change window, we compare observed ASI-related dissolution and formation with two volume-matched references: Random, which uniformly samples eligible changes, and RecSys, which represents recommendation-shaped network evolution. Our framework is general to recommender systems; here we instantiate RecSys with matrix factorization, a commonly used recommendation algorithm, to rank candidate new ties and likely-to-break ties. For each resulting graph, we run signed community detection and compute a toxicity-based disruption proxy $\Psi$, defined as the ratio of average across-subgroup toxicity to average within-subgroup toxicity. We report dissolution lift $R^-_n=\Psi_t^-/\Psi_n^-$, formation lift $R^+_n=\Psi_t^+/\Psi_n^+$, and normalized dominance $R_n=R^-_n/R^+_n$ for $n\in\{\text{Random},\text{RecSys}\}$.

\vspace{-1.5ex}
\subsection{Result Analysis and Insights}

\begin{table}[t]
\centering

\caption{Macro-level subgroup disruption under Definition I and II.
$R^-_n$ and $R^+_n$ denote dissolution and formation lift relative to null mechanism $n\in\{\mathrm{Random},\mathrm{RecSys}\}$. Random denotes uniform random rewiring and RecSys denotes the matrix-factorization recommendation null. The corresponding normalized dominance $R_n=R^-_n/R^+_n$ is visualized in Figure~\ref{fig:macro_ratio}.
Bluesky shows more persistent dissolution dominance, especially under structural-based ASI, while X is more sensitive to the time window and null mechanism.}
\label{tab:macro}
\small
\setlength{\tabcolsep}{4pt}
\renewcommand{\arraystretch}{1}
\begin{tabular}{@{}ccccccc@{}}
\toprule
\multirow{2}{*}{Dataset} & \multirow{2}{*}{Window} & \multirow{2}{*}{Def.}
& \multicolumn{2}{c}{Random null}
& \multicolumn{2}{c}{RecSys null} \\
\cmidrule(lr){4-5} \cmidrule(lr){6-7}
& & 
& $R^-_{\text{Random}}$ & $R^+_{\text{Random}}$
& $R^-_{\text{RecSys}}$ & $R^+_{\text{RecSys}}$ \\
\midrule
\multirow{6}{*}{X}
& \multirow{2}{*}{0--4w}
& I  & \mstd{1.13}{0.40} & \mstd{0.85}{0.62}
     & \mstd{2.44}{0.73} & \mstd{2.89}{4.43} \\
& 
& II & \mstd{0.40}{0.14} & \mstd{0.31}{0.40}
     & \mstd{0.77}{1.22} & \mstd{0.61}{0.25} \\
\cmidrule(l){2-7}
& \multirow{2}{*}{0--8w}
& I  & \mstd{0.32}{0.15} & \mstd{0.89}{0.49}
     & \mstd{1.05}{0.81} & \mstd{1.64}{1.80} \\
&
& II & \mstd{0.45}{0.26} & \mstd{0.56}{0.34}
     & \mstd{0.51}{0.62} & \mstd{1.24}{0.39} \\
\cmidrule(l){2-7}
& \multirow{2}{*}{0--12w}
& I  & \mstd{0.31}{0.10} & \mstd{0.69}{0.44}
     & \mstd{0.82}{0.71} & \mstd{1.65}{1.74} \\
&
& II & \mstd{0.73}{0.78} & \mstd{0.58}{0.33}
     & \mstd{1.09}{2.12} & \mstd{1.63}{2.61} \\
\midrule
\multirow{6}{*}{Bluesky}
& \multirow{2}{*}{0--8m}
& I  & \mstd{0.91}{0.81} & \mstd{0.43}{0.09}
     & \mstd{1.15}{0.18} & \mstd{0.63}{0.08} \\
&
& II & \mstd{1.47}{0.42} & \mstd{0.76}{0.35}
     & \mstd{1.67}{0.56} & \mstd{0.90}{0.41} \\
\cmidrule(l){2-7}
& \multirow{2}{*}{0--1.3y}
& I  & \mstd{0.94}{0.38} & \mstd{0.42}{0.08}
     & \mstd{1.18}{0.27} & \mstd{0.45}{0.05} \\
&
& II & \mstd{1.16}{0.46} & \mstd{0.83}{0.22}
     & \mstd{1.19}{0.50} & \mstd{0.97}{0.25} \\
\cmidrule(l){2-7}
& \multirow{2}{*}{0--2y}
& I  & \mstd{0.89}{0.29} & \mstd{0.53}{0.10}
     & \mstd{1.06}{0.35} & \mstd{0.55}{0.08} \\
&
& II & \mstd{1.08}{0.36} & \mstd{1.06}{0.24}
     & \mstd{1.05}{0.44} & \mstd{1.12}{0.26} \\
\bottomrule
\end{tabular}
\vspace{-2ex}
\end{table}

\begin{figure}[t]
\centering
\includegraphics[width=0.48\textwidth]{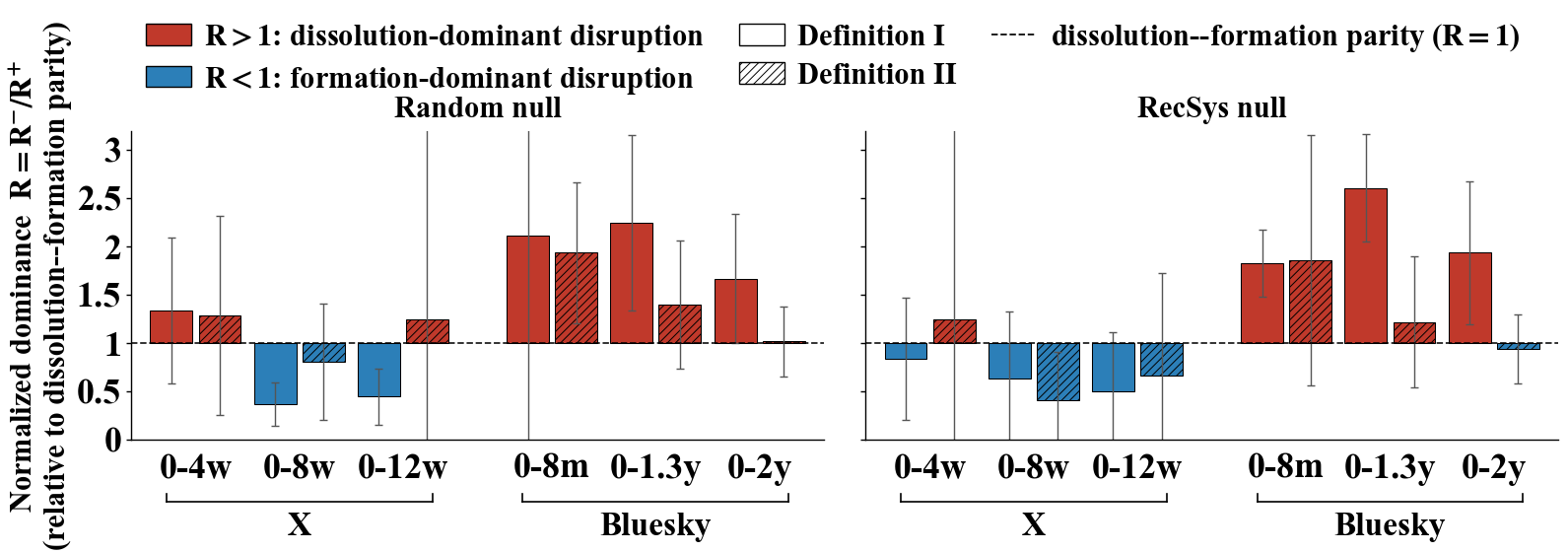}
\vspace{-3.75ex}
\caption{Normalized dominance $R = R^-/R^+$ across datasets, windows, and ASI definitions under the Random and RecSys
reference mechanisms. Bars diverge from the equality baseline $R=1$ (dashed): red (up) marks $R>1$ (dissolution dominant), blue (down) marks $R<1$ (formation dominant); solid = Def.~I, hatched = Def.~II. Error bars: $\pm 1$ SD.}\label{fig:macro_ratio}
\vskip -3.5ex
\end{figure}

Table~\ref{tab:macro} reports dissolution and formation lifts under the Random and RecSys reference mechanisms, and Figure~\ref{fig:macro_ratio} visualizes the corresponding normalized dominance $R=R^{-}/R^{+}$. We interpret these results by considering both the normalized dominance and the two underlying lift values, since $R>1$ can arise either from strong dissolution lift or from weak formation lift.

\noindent\textbf{Platform context shapes the macro-level pathway of ASI.}
Bluesky shows a more persistent dissolution-dominant pattern than X. Under Definition I, Bluesky has $R>1$ across all windows and under both reference mechanisms, indicating that structural-based ASI is more strongly associated with subgroup disruption through dissolution than through formation. This pattern is especially clear under the RecSys reference, where $R^{-}_{\mathrm{RecSys}}>1$ and $R^{+}_{\mathrm{RecSys}}<1$ in all three windows. X shows a less stable pattern. The early 0--4w window has some dissolution dominance under the Random null, but the signal weakens or reverses in longer windows. Under the RecSys reference, X is often formation-dominant, especially for Definition I. Thus, Bluesky shows a more stable dissolution pathway compared with X.

\vspace{1.5ex}
\noindent\textbf{The two ASI definitions reveal different disruption signals.}
Definition I, based on structural-based disconnection, provides the clearest macro-level signal of subgroup separation. On Bluesky, structural-based ASI remains dissolution-dominant from the short to the long window, suggesting that observed disconnections consistently align with the separation of cohesive subgroups. Definition II, based on content toxicity, also shows disruption, but its signal is more mixed. On Bluesky, dissolution lift under Definition II is above one across windows, showing that toxic directed communication is associated with subgroup disruption. However, formation lift increases over time, making the long-horizon pattern less purely dissolution-driven. On X, Definition II is weaker and more reference-dependent. Overall, structural-based negativity gives a more persistent signal of cohesive subgroup disruption, while content-based toxicity captures a broader and more time-varying form of conflict.

\vspace{1ex}
\noindent\textbf{Temporal evolution distinguishes persistent separation from mixed disruption.}
The time-window comparison shows how ASI-related disruption evolves. On Bluesky, structural-based ASI maintains dissolution dominance from 0--8m to 0--2y. This persistence is not simply due to a growing dissolution lift; under the Random null, dissolution lift is close to one, while formation lift remains much lower. Under the RecSys reference, the pattern is stronger because observed dissolution exceeds the recommendation-based reference and observed formation remains below it. Content-based ASI evolves differently. Under Definition II, Bluesky begins with strong dissolution dominance, but the dominance weakens by 0--2y as formation lift becomes comparable to dissolution lift. This suggests that toxic communication can mark early subgroup separation, but over longer periods, it also appears in formation disruption. On X, the early dissolution signal is not stable over longer windows, indicating a more transient or context-dependent pattern.

\vspace{1ex}
\noindent\textbf{Random and RecSys separate churn from recommendation-shaped rewiring.}
The Random null asks whether observed ASI-related changes are more disruptive than uniformly sampled eligible changes with the same change volume. It controls for disruption caused by network churn alone. The RecSys reference asks a different question: whether observed changes are more or less disruptive than changes selected by a matrix-factorization-based recommendation mechanism. RecSys is therefore an algorithmic reference, not a pure null model. It helps assess whether the observed disruption is specific to user behavior or whether recommendation-shaped rewiring could produce a different disruption profile. This distinction is important: Bluesky structural-based ASI remains dissolution-dominant under both references, while X changes substantially when moving from Random to RecSys.

\vspace{1ex}
\noindent\textbf{Formation-related disruption is also part of ASI-driven network evolution.}
The results also show that subgroup disruption cannot be explained only by dissolution. On X, and for content-based ASI over longer Bluesky windows, formation lift becomes important and sometimes exceeds dissolution lift. This means that newly formed ties can also increase the disruption proxy, especially when they connect users or communities in ways associated with higher cross-subgroup toxicity. This is important for interpreting ASI evolution: adverse dynamics may disrupt cohesive subgroups not only by breaking existing relations, but also by shaping which new relations emerge around conflict. The RecSys reference is useful here because it indicates whether recommendation-like rewiring would attenuate or amplify this formation-related disruption relative to observed behavior.

\vspace{0.25ex}
\noindent\textbf{Overall, structural-based ASI gives the most persistent evidence of cohesive subgroup disruption.}
Across datasets, windows, and reference mechanisms, Definition I provides the strongest and most stable evidence that ASI contributes to subgroup disruption through dissolution, especially on Bluesky. Definition II provides complementary evidence that toxic communication is also related to subgroup disruption, but its effect is more time-dependent and more entangled with formation. These results support the main macro-level conclusion: structural- and content-based ASI both affect the evolution of online communities, but structural disconnection more persistently marks the disruption of cohesive subgroups, while content toxicity captures a broader conflict process that can operate through both dissolution and formation.
\section{Discussion}\label{sec:discuss}

Our results show that ASI is not a single-scale phenomenon. At the micro level, adverse ties are consistently more associated with higher-toxicity interactions than ordinary friend ties, while content-based ASI appears both within and across community boundaries. This suggests that ASI reflects both internal community friction and boundary-spanning conflict. At the meso level, matched analyses show that exposure to adverse actions is associated with subsequent adverse behavior among neighboring users, with stronger effects for structural disconnection and for stronger exposed--source ties. At the macro level, ASI-related tie evolution can disrupt cohesive subgroups, but the disruption pathway depends on both the ASI definition and the reference mechanism. Structural ASI provides the most persistent signal of dissolution-driven subgroup disruption, especially on Bluesky, while content-based ASI captures a broader and more time-varying conflict process. Together, these findings suggest that structural and content-based ASI are complementary signals: one captures relational breakdown, while the other captures communicative hostility and its local context.

\section{Limitation and Future Work}\label{sec:limitations}

This study has several limitations. First, our meso-level analysis relies on observational matching, so unobserved confounding and platform-specific exposure mechanisms may still affect the estimated peer influence. Second, X and Bluesky differ in observation windows, available interaction types, and structural signals, which limits direct causal comparison across platforms. Third, toxicity scores and subgroup-disruption proxies cannot fully capture user intent or conversational context. Future work will extend the framework to more platforms, incorporate richer behavioral and content signals, and study how recommendation systems may amplify or mitigate ASI-driven network evolution.

\section{Conclusion}\label{sec:con}

In this paper, we study ASIs through a unified set of definitions and a multi-level evaluation framework on two platforms with different interaction structures.
Our micro-level results characterize how negative behaviors concentrate within and across communities and vary with user prominence.
The meso-level results quantify local influence by estimating how exposure to negative acts changes the future negative act propensity of neighbors.
The macro-level case study connects tie evolution to cohesive subgroup disruption and shows that the inferred disruption depends on both the ASI definition and the counterfactual mechanism.

\appendix
\section*{Appendix}

\section{Spectral Rationale for Negative-Edge Reweighting}
\label{sec:app_negative_weight}

This section provides the rationale for the negative-edge weighting used in our signed community detection. In our datasets, positive social ties are abundant and often long-lived, while ASI edges are much sparser and usually capture more recent conflict signals. If positive and negative edges are weighted equally, the clustering objective can be dominated by the positive follow graph, making the partition less sensitive to rare but informative adverse relations. We therefore use a fixed negative-edge weight $\alpha>1$ in the main analysis, with $\alpha=10$ unless otherwise stated. This value is fixed before downstream analysis and is not tuned to optimize any reported outcome. Its purpose is to make sparse negative signals visible in the structural partition. This design is also consistent with prior evidence that negative interactions can have stronger social effects than positive ones \cite{baumeister2001bad,rozin2001negativity}.

\paragraph{Signed spectral clustering.}
Let $A^+$ and $A^-$ denote the symmetrized positive and negative adjacency matrices. Positive edges correspond to social ties, and negative edges correspond to ASI ties under the chosen ASI definition. For a negative-edge weight $\alpha>0$, we define the signed adjacency matrix $S_\alpha = A^+ - \alpha A^-$ and the signed degree matrix $D_\alpha = \mathrm{diag}\big((A^+ + \alpha A^-)\mathbf{1}\big)$. We then use the symmetric normalized signed Laplacian $L_\alpha = I - D_\alpha^{-1/2} S_\alpha D_\alpha^{-1/2}$ and run spectral clustering on the bottom-$k$ eigenvectors of $L_\alpha$ \cite{kunegis2010spectral,chiang2011exploiting}. This objective favors partitions where positive edges remain within communities and negative edges fall across communities, following the intuition of structural balance theory \cite{cartwright1956structural,heider1946attitudes}. For a two-way relaxed embedding $z$, the corresponding quadratic form is
\[
z^\top (D_\alpha-S_\alpha) z
=
\sum_{i<j} A^+_{ij}(z_i-z_j)^2
+
\alpha\sum_{i<j} A^-_{ij}(z_i+z_j)^2.
\]
The first term penalizes separating endpoints of positive edges, and the second term penalizes placing endpoints of negative edges on the same side of the partition. Increasing $\alpha$ hence directly strengthens the influence of negative edges in the partitioning objective.

\paragraph{Stability under graph updates.}
The effect of network growth and edge updates can be understood through standard spectral perturbation. Let $L_\alpha$ and $\widetilde L_\alpha=L_\alpha+E$ be the signed Laplacians of two consecutive aligned snapshots, and let $U_k$ and $\widetilde U_k$ denote their bottom-$k$ eigenspaces. If the eigengap $\delta_k=\lambda_{k+1}(L_\alpha)-\lambda_k(L_\alpha)$
is positive and $\|E\|_2\leq \varepsilon$, the Davis--Kahan theorem gives $\|\widetilde U_k\widetilde U_k^\top-U_kU_k^\top\|_F
\leq
\frac{2\sqrt{2}\varepsilon}{\delta_k}$. Thus, when the eigengap is sufficiently large, small changes in the signed graph lead only to bounded movement in the spectral embedding. This supports our use of signed clustering across temporal snapshots: persistent friend--foe patterns are unlikely to be caused by minor edge perturbations alone.

\paragraph{Effect of the negative-edge weight.}
The weight $\alpha$ determines how strongly negative edges affect the signed Laplacian. Since $L_\alpha$ changes continuously with $\alpha$, small changes in $\alpha$ have limited effect when the eigengap is large. When negative edges form coherent conflict structure, however, increasing $\alpha$ can shift the embedding and reveal partitions that would otherwise be dominated by the dense positive follow graph. This motivates our fixed choice of $\alpha=10$ as it keeps sparse ASI signals visible in community detection.

\section{Dataset Statistics}

Bluesky covers a longer period, from April 2023 to May 2025, with 17,017 tracked users, 15.4M follow edges, 1.78M block events, and 10.47M posts. X covers a shorter 16-week window from April 7 to July 28, 2022, but is larger in scale, with 127,829 tracked users, 90.95M follow edges, 4.69M block/unfollow events, and 30.34M tweets. The datasets also differ in interaction types: Bluesky includes replies, mentions, and quotes, while X includes replies, mentions, retweets, and original tweets. These motivate our emphasis on within-platform trends, with cross-platform comparisons used to identify qualitative similarities and differences in ASI dynamics.

\begingroup
\setlength{\textfloatsep}{2pt plus 1pt minus 1pt} 

\begin{algorithm}[t]
\caption{Pair construction via replacement-based matching at exposure time $T^\star$.}
\label{alg:pair-construction}
\small
\DontPrintSemicolon
\KwIn{Treated triads $\mathcal{T}$; windows $(\Delta,W_{\mathrm{pre}},W_{\mathrm{out}},w)$; encoder \textsc{Enc}; constraints $(\mathcal{C}_e,\mathcal{C}_t)$; similarity $\mathrm{sim}$.}
\KwOut{Matched outcome pairs for Match-$U_t$ and Match-$U_e$.}

\SetKwFunction{BestMatch}{BestMatch}
\SetKwProg{Fn}{Function}{:}{}
\Fn{\BestMatch{$\mathcal{P}, z_{\mathrm{ref}}$}}{
  \Return{$\arg\max_{u\in\mathcal{P}} \mathrm{sim}(z_{\mathrm{ref}}, z_u)$}\;
}

\ForEach{$(U_e,U_s,U_t)\in\mathcal{T}$}{
  Determine $T^\star$\;
  \If{\textsc{InsufficientPosts}$(\cdot)$}{\textbf{continue}\;}
  $z_e\!\gets\!\textsc{Enc}(U_e)$;\; $z_t\!\gets\!\textsc{Enc}(U_t)$\;
  $y_T\!\gets\!Y_T(U_e,U_s,U_t)$ \tcp*[r]{in $(T^\star,T^\star+W_{\mathrm{out}}]$}

  \tcp{Match-$U_t$ (replace target; hold $U_e,U_s$)}
  $\mathcal{P}_t\!\gets\!\textsc{CandTargets}(U_e,U_s,U_t,T^\star;\mathcal{C}_t)$\;
  \If{$\mathcal{P}_t\neq\emptyset$}{
    $U_t^{\mathrm{ctrl}}\!\gets\!\BestMatch(\mathcal{P}_t,z_t)$\;
    Store $(y_T,\;Y_C(U_e,U_s,U_t^{\mathrm{ctrl}}))$\;
  }

  \tcp{Match-$U_e$ (replace exposed; hold $U_s,U_t$)}
  $\mathcal{P}_e\!\gets\!\textsc{CandExposed}(U_e,U_s,U_t,T^\star;\mathcal{C}_e)$\;
  \If{$\mathcal{P}_e\neq\emptyset$}{
    $U_e^{\mathrm{ctrl}}\!\gets\!\BestMatch(\mathcal{P}_e,z_e)$\;
    Store $(y_T,\;Y_C(U_e^{\mathrm{ctrl}},U_s,U_t))$\;
  }
}
Aggregate $(a,b,c,d)$ and compute $\hat{\tau}$, McNemar statistics, and CIs as in \S\ref{sec:estimator}.\;
\end{algorithm}
\endgroup

\section{Pair Construction Algorithm}
\label{sec:algo-pair}
Algorithm~\ref{alg:pair-construction} formalizes the replacement-based matching procedure described in §\ref{sec:treat-control}, taking treated triads $(U_e,U_s,U_t)$ at exposure time $T^\star$ and returning the treated–control outcome pairs aggregated by the McNemar test in §\ref{sec:estimator}. For each triad, we encode user histories with \textsc{Enc}, draw admissible candidates under $(\mathcal{C}_t,\mathcal{C}_e)$, select the most similar control via cosine similarity, and store the resulting pair $(y_T, Y_C)$ separately for the Match-$U_t$ and Match-$U_e$ designs.

\begin{table}[t]
\centering
\tiny 
\caption{Bluesky blocking outcome: matched treatment and control rates per 1k interactions, with multiplicative T/C ratios and McNemar exact $p$ values.}
\label{tab:bluesky_original}
\setlength{\tabcolsep}{1pt}
\renewcommand{\arraystretch}{1.08}

\begin{tabular}{@{}cc
@{\hspace{3pt}}|@{\hspace{3pt}}rrrc
@{\hspace{4pt}}|@{\hspace{3pt}}rrrc
@{\hspace{4pt}}|@{\hspace{3pt}}rrrc@{}}
\toprule
\multirow{2}{*}{\makecell[c]{Match\\Set.}}
& \multirow{2}{*}{\makecell[c]{Interact.\\Type}}
& \multicolumn{4}{c}{S$\to$S}
& \multicolumn{4}{c}{C$\to$S}
& \multicolumn{4}{c}{C$\to$C} \\
\cmidrule(lr){3-6} \cmidrule(lr){7-10} \cmidrule(lr){11-14}
&
& Tr./1k & Cr./1k & $\times$ & $p$
& Tr./1k & Cr./1k & $\times$ & $p$
& Tr./1k & Cr./1k & $\times$ & $p$ \\
\midrule

\multirow{4}{*}{$U_{\mathrm{t}}'$}
& O
& 1.118 & 0.046 & 24.3 & $4.44\mathrm{e}{-16}$
& 0.175 & 0.048 & 3.6 & $<1\mathrm{e}{-10}$
& 2.803 & 3.687 & 0.8 & $4.46\mathrm{e}{-119}$ \\
& M
& 5.889 & 0.000 & $\infty$ & $6.10\mathrm{e}{-5}$
& 0.131 & 0.026 & 5.0 & $0.219$
& 2.053 & 5.079 & 0.4 & $1.11\mathrm{e}{-12}$ \\
& Q
& 2.173 & 0.000 & $\infty$ & $<1\mathrm{e}{-10}$
& 0.151 & 0.046 & 3.3 & $<1\mathrm{e}{-10}$
& 2.952 & 3.661 & 0.8 & $6.56\mathrm{e}{-44}$ \\
& R
& 0.736 & 0.059 & 12.5 & $1.12\mathrm{e}{-18}$
& 0.207 & 0.051 & 4.1 & $<1\mathrm{e}{-10}$
& 2.619 & 3.693 & 0.7 & $3.78\mathrm{e}{-77}$ \\

\midrule

\multirow{4}{*}{$U_{\mathrm{t}}''$}
& O
& 1.107 & 0.038 & 29.1 & $<1\mathrm{e}{-10}$
& 0.165 & 0.118 & 1.4 & $4.34\mathrm{e}{-10}$
& 3.059 & 2.369 & 1.3 & $1.47\mathrm{e}{-100}$ \\
& M
& 4.897 & 0.000 & $\infty$ & $6.10\mathrm{e}{-5}$
& 0.105 & 0.105 & 1.0 & $1.000$
& 1.803 & 2.830 & 0.6 & $9.61\mathrm{e}{-4}$ \\
& Q
& 2.171 & 0.000 & $\infty$ & $<1\mathrm{e}{-10}$
& 0.158 & 0.111 & 1.4 & $4.62\mathrm{e}{-6}$
& 3.406 & 2.554 & 1.3 & $8.45\mathrm{e}{-72}$ \\
& R
& 0.722 & 0.049 & 14.7 & $<1\mathrm{e}{-10}$
& 0.174 & 0.126 & 1.4 & $2.20\mathrm{e}{-5}$
& 2.717 & 2.164 & 1.3 & $1.25\mathrm{e}{-35}$ \\

\midrule

\multirow{4}{*}{$U_{\mathrm{e}}'$}
& O
& 1.032 & 0.299 & 3.5 & $5.40\mathrm{e}{-16}$
& 0.163 & 0.101 & 1.6 & $2.46\mathrm{e}{-18}$
& 2.989 & 1.493 & 2.0 & $<1\mathrm{e}{-10}$ \\
& M
& 4.912 & 0.655 & 7.5 & $2.40\mathrm{e}{-3}$
& 0.107 & 0.085 & 1.3 & $1.000$
& 1.923 & 1.004 & 1.9 & $2.42\mathrm{e}{-4}$ \\
& Q
& 1.928 & 0.429 & 4.5 & $1.40\mathrm{e}{-8}$
& 0.154 & 0.082 & 1.9 & $5.88\mathrm{e}{-14}$
& 3.390 & 1.989 & 1.7 & $1.25\mathrm{e}{-217}$ \\
& R
& 0.693 & 0.256 & 2.7 & $8.00\mathrm{e}{-7}$
& 0.173 & 0.120 & 1.4 & $7.98\mathrm{e}{-7}$
& 2.592 & 0.985 & 2.6 & $<1\mathrm{e}{-10}$ \\

\bottomrule
\end{tabular}
\begin{minipage}{\columnwidth}
\vspace{2ex}
\footnotesize{\hspace{2ex} Note: O: Overall; M: Mention; Q: Quote; R: Reply.}
\end{minipage}
\vskip -1ex
\end{table}

\vspace{-2ex}
\begin{table}[t]
\centering
\small
\setlength{\tabcolsep}{2pt}
\caption{X unfollowing outcome: matched Treatment vs.\ Control rates per 1k interactions.
Diff is multiplicative (T/C). McNemar exact $p$ shown. 
} 
\label{tab:X_original}
\renewcommand{\arraystretch}{1.05}

\begin{tabular}{@{}cccccccccc@{}}
\toprule
\multirow{2}{*}{\makecell[c]{Match\\Set.}}
& \multirow{2}{*}{\makecell[c]{Temporal\\Group}}
& \multicolumn{4}{c}{C$\to$S}
& \multicolumn{4}{c}{C$\to$C} \\
\cmidrule(lr){3-6} \cmidrule(lr){7-10}
& 
& Tr./1k & Cr./1k & $\times$ & $p$
& Tr./1k & Cr./1k & $\times$ & $p$ \\
\midrule

\multirow{3}{*}{$U_{\mathrm{t}}'$}
& $W{=}1$ 
& 0.156 & 0.312 & 0.5 & \pval{$5.08\mathrm{e}{-1}$}
& 3.769 & 5.270 & 0.7 & \pval{$4.52\mathrm{e}{-3}$} \\
& $W{=}2$ 
& 0.364 & 0.468 & 0.8 & \pval{$8.04\mathrm{e}{-1}$}
& 6.435 & 9.836 & 0.7 & \pval{$8.07\mathrm{e}{-7}$} \\
& $W{=}3$ 
& 0.676 & 0.676 & 1.0 & \pval{$1.000$}
& 9.438 & 13.115 & 0.7 & \pval{$4.18\mathrm{e}{-6}$} \\

\addlinespace[0.35em]
\midrule

\multirow{3}{*}{$U_{\mathrm{t}}''$}
& $W{=}1$ 
& 0.587 & 0.130 & 4.5 & \pval{$2.58\mathrm{e}{-3}$}
& 5.698 & 5.169 & 1.1 & \pval{$0.379$} \\
& $W{=}2$ 
& 1.075 & 0.456 & 2.4 & \pval{$6.61\mathrm{e}{-3}$}
& 10.214 & 8.968 & 1.1 & \pval{$0.103$} \\
& $W{=}3$ 
& 1.499 & 0.619 & 2.4 & \pval{$8.98\mathrm{e}{-4}$}
& 15.134 & 11.179 & 1.4 & \pval{$4.84\mathrm{e}{-6}$} \\

\addlinespace[0.35em]
\midrule

\multirow{3}{*}{$U_{\mathrm{e}}'$}
& $W{=}1$ 
& 0.499 & 0.499 & 1.0 & \pval{$1.000$}
& 3.794 & 1.417 & 2.7 & \pval{$1.18\mathrm{e}{-6}$} \\
& $W{=}2$ 
& 0.940 & 1.028 & 0.9 & \pval{$8.04\mathrm{e}{-1}$}
& 7.359 & 2.102 & 3.5 & \pval{$3.75\mathrm{e}{-16}$} \\
& $W{=}3$ 
& 1.352 & 1.410 & 1.0 & \pval{$9.16\mathrm{e}{-1}$}
& 11.518 & 2.697 & 4.3 & \pval{$1.48\mathrm{e}{-29}$} \\

\bottomrule
\end{tabular}
\vspace{-4ex}
\end{table}

\section{Additional Results}\label{sec:appendix_add}

Tables~\ref{tab:bluesky_original} and \ref{tab:X_original} report the
full per-interaction-type and per-window breakdowns underlying the
summary in §\ref{sec:meso-results}. We highlight two patterns that reinforce the
main finding.

\vspace{1ex}
\noindent\textbf{The headline cascade signals are robust to the
breakdown.} On Bluesky, the S$\to$S coupling produces strongly
positive paired contrasts under all three matching designs and across
interaction types, and the C$\to$C coupling under Match-$U_e'$ remains
positive across modalities. On X, the C$\to$C coupling under
Match-$U_e'$ is significant in every temporal window. The main-text
effects are not driven by any single interaction subtype or
outcome horizon.

\vspace{1ex}
\noindent\textbf{The three replacement designs probe complementary
counterfactuals.} Match-$U_t'$ and Match-$U_t''$ rows differ in
systematic ways across the content couplings, reflecting the distinct
composition of their control targets: near-$U_s$ controls share more
of the baseline interaction context with the treated target, while
far-from-$U_s$ controls remove that shared context. Match-$U_e'$
holds the source--target pair fixed and is the direct
test of follow-tie-mediated transmission, which we use as the primary
reference in §\ref{sec:meso-results}.

\bibliographystyle{ACM-Reference-Format}
\bibliography{ref}

@article{rubin2005causal,
  title={Causal inference using potential outcomes: Design, modeling, decisions},
  author={Rubin, Donald B},
  journal={Journal of the American statistical Association},
  volume={100},
  number={469},
  pages={322--331},
  year={2005},
  publisher={Taylor \& Francis}
}

@article{rosenbaum1983central,
  title={The central role of the propensity score in observational studies for causal effects},
  author={Rosenbaum, Paul R and Rubin, Donald B},
  journal={Biometrika},
  volume={70},
  number={1},
  pages={41--55},
  year={1983},
  publisher={Oxford University Press}
}

@inproceedings{russo2024stranger,
  title={Stranger danger! cross-community interactions with fringe users increase the growth of fringe communities on reddit},
  author={Russo, Giuseppe and Ribeiro, Manoel Horta and West, Robert},
  booktitle={Proceedings of the International AAAI Conference on Web and Social Media},
  volume={18},
  pages={1342--1353},
  year={2024}
}

@book{rosenbaum2010design,
  title={Design of observational studies},
  author={Rosenbaum, Paul R and Rosenbaum, P and Briskman},
  volume={10},
  year={2010},
  publisher={Springer}
}

@article{stuart2010matching,
  title={Matching methods for causal inference: A review and a look forward},
  author={Stuart, Elizabeth A},
  journal={Statistical science: a review journal of the Institute of Mathematical Statistics},
  volume={25},
  number={1},
  pages={1},
  year={2010}
}

@article{li2001balanced,
  title={Balanced risk set matching},
  author={Li, Yunfei Paul and Propert, Kathleen J and Rosenbaum, Paul R},
  journal={Journal of the American Statistical Association},
  volume={96},
  number={455},
  pages={870--882},
  year={2001},
  publisher={Taylor \& Francis}
}

@article{hudgens2008toward,
  title={Toward causal inference with interference},
  author={Hudgens, Michael G and Halloran, M Elizabeth},
  journal={Journal of the american statistical association},
  volume={103},
  number={482},
  pages={832--842},
  year={2008},
  publisher={Taylor \& Francis}
}

@article{reimers2019sentence,
  title={Sentence-bert: Sentence embeddings using siamese bert-networks},
  author={Reimers, Nils and Gurevych, Iryna},
  journal={arXiv preprint arXiv:1908.10084},
  year={2019}
}

@inproceedings{veitch2020adapting,
  title={Adapting text embeddings for causal inference},
  author={Veitch, Victor and Sridhar, Dhanya and Blei, David},
  booktitle={Conference on uncertainty in artificial intelligence},
  pages={919--928},
  year={2020},
  organization={PMLR}
}

@article{mcnemar1947note,
  title={Note on the sampling error of the difference between correlated proportions or percentages},
  author={McNemar, Quinn},
  journal={Psychometrika},
  volume={12},
  number={2},
  pages={153--157},
  year={1947},
  publisher={Springer-Verlag}
}

@article{sasahara2021social,
  title={Social influence and unfollowing accelerate the emergence of echo chambers},
  author={Sasahara, Kazutoshi and Chen, Wen and Peng, Hao and Ciampaglia, Giovanni Luca and Flammini, Alessandro and Menczer, Filippo},
  journal={Journal of Computational Social Science},
  volume={4},
  number={1},
  pages={381--402},
  year={2021},
  publisher={Springer}
}

@article{tong2017efficient,
  title={An efficient randomized algorithm for rumor blocking in online social networks},
  author={Tong, Guangmo and Wu, Weili and Guo, Ling and Li, Deying and Liu, Cong and Liu, Bin and Du, Ding-Zhu},
  journal={IEEE Transactions on Network Science and Engineering},
  volume={7},
  number={2},
  pages={845--854},
  year={2017},
  publisher={IEEE}
}

@article{pham2020multi,
  title={Multi-topic misinformation blocking with budget constraint on online social networks},
  author={Pham, Dung V and Nguyen, Giang L and Nguyen, Tu N and Pham, Canh V and Nguyen, Anh V},
  journal={IEEE Access},
  volume={8},
  pages={78879--78889},
  year={2020},
  publisher={IEEE}
}

@inproceedings{dey2017centrality,
  title={Centrality based information blocking and influence minimization in online social network},
  author={Dey, Paramita and Roy, Sarbani},
  booktitle={2017 IEEE international conference on advanced networks and telecommunications systems (ANTS)},
  pages={1--6},
  year={2017},
  organization={IEEE}
}

@inproceedings{wu2020mining,
  title={Mining unfollow behavior in large-scale online social networks via spatial-temporal interaction},
  author={Wu, Haozhe and Hu, Zhiyuan and Jia, Jia and Bu, Yaohua and He, Xiangnan and Chua, Tat-Seng},
  booktitle={Proceedings of the AAAI Conference on Artificial Intelligence},
  volume={34},
  number={01},
  pages={254--261},
  year={2020}
}

@inproceedings{kwak2011fragile,
  title={Fragile online relationship: a first look at unfollow dynamics in twitter},
  author={Kwak, Haewoon and Chun, Hyunwoo and Moon, Sue},
  booktitle={Proceedings of the SIGCHI conference on human factors in computing systems},
  pages={1091--1100},
  year={2011}
}

@inproceedings{kivran2011impact,
  title={The impact of network structure on breaking ties in online social networks: unfollowing on twitter},
  author={Kivran-Swaine, Funda and Govindan, Priya and Naaman, Mor},
  booktitle={Proceedings of the SIGCHI conference on human factors in computing systems},
  pages={1101--1104},
  year={2011}
}

@article{twenge2013does,
  title={Does online social media lead to social connection or social disconnection?},
  author={Twenge, Jean M},
  journal={Journal of College and Character},
  volume={14},
  number={1},
  pages={11--20},
  year={2013},
  publisher={Taylor \& Francis}
}

@article{jorge2019social,
  title={Social media, interrupted: Users recounting temporary disconnection on Instagram},
  author={Jorge, Ana},
  journal={Social Media+ Society},
  volume={5},
  number={4},
  pages={2056305119881691},
  year={2019},
  publisher={SAGE Publications Sage UK: London, England}
}

@article{sheth2022defining,
  title={Defining and detecting toxicity on social media: context and knowledge are key},
  author={Sheth, Amit and Shalin, Valerie L and Kursuncu, Ugur},
  journal={Neurocomputing},
  volume={490},
  pages={312--318},
  year={2022},
  publisher={Elsevier}
}

@inproceedings{ali2021understanding,
  title={Understanding the effect of deplatforming on social networks},
  author={Ali, Shiza and Saeed, Mohammad Hammas and Aldreabi, Esraa and Blackburn, Jeremy and De Cristofaro, Emiliano and Zannettou, Savvas and Stringhini, Gianluca},
  booktitle={Proceedings of the 13th ACM Web Science Conference 2021},
  pages={187--195},
  year={2021}
}

@article{harrigan2020negative,
  title={Negative ties and signed graphs research: Stimulating research on dissociative forces in social networks},
  author={Harrigan, Nicholas M and Labianca, Giuseppe Joe and Agneessens, Filip},
  journal={Social Networks},
  volume={60},
  pages={1--10},
  year={2020},
  publisher={Elsevier}
}

@article{sonnemans2006dynamics,
  title={On the dynamics of social ties structures in groups},
  author={Sonnemans, Joep and Van Dijk, Frans and Van Winden, Frans},
  journal={Journal of economic psychology},
  volume={27},
  number={2},
  pages={187--204},
  year={2006},
  publisher={Elsevier}
}

@article{cote2019evolution,
  title={The Evolution of Social Network Theory: Perceived Impact on Developing Networking Relationships.},
  author={Cote, Robert},
  journal={American Journal of Management},
  volume={19},
  number={3},
  year={2019}
}

@article{candellone2025negative,
  title={Negative Ties Highlight Hidden Extremes in Social Media Polarization},
  author={Candellone, Elena and Babul, Shazia'Ayn and Togay, {\"O}zg{\"u}r and Bovet, Alexandre and Garcia-Bernardo, Javier},
  journal={arXiv preprint arXiv:2501.05590},
  year={2025}
}

@article{marlowe2017digital,
  title={Digital belongings: The intersections of social cohesion, connectivity and digital media},
  author={Marlowe, Jay M and Bartley, Allen and Collins, Francis},
  journal={Ethnicities},
  volume={17},
  number={1},
  pages={85--102},
  year={2017},
  publisher={SAGE Publications Sage UK: London, England}
}

@article{kwak2017understanding,
  title={Understanding the process of social network evolution: Online-offline integrated analysis of social tie formation},
  author={Kwak, Doyeon and Kim, Wonjoon},
  journal={Plos one},
  volume={12},
  number={5},
  pages={e0177729},
  year={2017},
  publisher={Public Library of Science San Francisco, CA USA}
}

@article{offer2021negative,
  title={Negative social ties: Prevalence and consequences},
  author={Offer, Shira},
  journal={Annual Review of Sociology},
  volume={47},
  number={1},
  pages={177--196},
  year={2021},
  publisher={Annual Reviews}
}

@article{shen2014evolution,
  title={The evolution of social ties online: A longitudinal study in a massively multiplayer online game},
  author={Shen, Cuihua and Monge, Peter and Williams, Dmitri},
  journal={Journal of the Association for Information Science and Technology},
  volume={65},
  number={10},
  pages={2127--2137},
  year={2014},
  publisher={Wiley Online Library}
}

@article{ganley2009ties,
  title={The ties that bind: Social network principles in online communities},
  author={Ganley, Dale and Lampe, Cliff},
  journal={Decision support systems},
  volume={47},
  number={3},
  pages={266--274},
  year={2009},
  publisher={Elsevier}
}

@inproceedings{leskovec2010predicting,
  title={Predicting positive and negative links in online social networks},
  author={Leskovec, Jure and Huttenlocher, Daniel and Kleinberg, Jon},
  booktitle={Proceedings of the 19th international conference on World wide web},
  pages={641--650},
  year={2010}
}

@article{esmailian2014mesoscopic,
  title={Mesoscopic analysis of online social networks-the role of negative ties},
  author={Esmailian, Pouya and Abtahi, Seyed Ebrahim and Jalili, Mahdi},
  journal={arXiv preprint arXiv:1411.6057},
  year={2014}
}

@article{bolibar2016macro,
  title={Macro, meso, micro: Broadening the ‘social’of social network analysis with a mixed methods approach},
  author={Bol{\'\i}bar, Mireia},
  journal={Quality \& Quantity},
  volume={50},
  number={5},
  pages={2217--2236},
  year={2016},
  publisher={Springer}
}

@article{liu2022weak,
  title={Weak ties matter: Social network dynamics of mobile media multiplexity and their impact on the social support and psychological well-being experienced by migrant workers},
  author={Liu, Piper Liping and Yeo, Tien Ee Dominic},
  journal={Mobile Media \& Communication},
  volume={10},
  number={1},
  pages={76--96},
  year={2022},
  publisher={SAGE Publications Sage UK: London, England}
}

@article{cinelli2021echo,
  title={The echo chamber effect on social media},
  author={Cinelli, Matteo and De Francisci Morales, Gianmarco and Galeazzi, Alessandro and Quattrociocchi, Walter and Starnini, Michele},
  journal={Proceedings of the national academy of sciences},
  volume={118},
  number={9},
  pages={e2023301118},
  year={2021},
  publisher={National Academy of Sciences}
}

@inproceedings{pedersen2019analyzing,
  title={Analyzing echo chambers: A logic of strong and weak ties},
  author={Pedersen, Mina Young and Smets, Sonja and {\AA}gotnes, Thomas},
  booktitle={International Workshop on Logic, Rationality and Interaction},
  pages={183--198},
  year={2019},
  organization={Springer}
}

@article{felmlee2016toxic,
  title={Toxic ties: Networks of friendship, dating, and cyber victimization},
  author={Felmlee, Diane and Faris, Robert},
  journal={Social psychology quarterly},
  volume={79},
  number={3},
  pages={243--262},
  year={2016},
  publisher={Sage Publications Sage CA: Los Angeles, CA}
}

@inproceedings{budak2011limiting,
  title={Limiting the spread of misinformation in social networks},
  author={Budak, Ceren and Agrawal, Divyakant and El Abbadi, Amr},
  booktitle={Proceedings of the 20th international conference on World wide web},
  pages={665--674},
  year={2011}
}

@article{huang2018will,
  title={Will triadic closure strengthen ties in social networks?},
  author={Huang, Hong and Dong, Yuxiao and Tang, Jie and Yang, Hongxia and Chawla, Nitesh V and Fu, Xiaoming},
  journal={ACM Transactions on Knowledge Discovery from Data (TKDD)},
  volume={12},
  number={3},
  pages={1--25},
  year={2018},
  publisher={ACM New York, NY, USA}
}

@article{song2019triadic,
  title={Triadic closure, homophily, and reciprocation: an empirical investigation of social ties between content providers},
  author={Song, Tingting and Tang, Qian and Huang, Jinghua},
  journal={Information Systems Research},
  volume={30},
  number={3},
  pages={912--926},
  year={2019},
  publisher={INFORMS}
}

@article{li2023information,
  title={Information cascades blocking through influential nodes identification on social networks},
  author={Li, Li and Zheng, Xiaohua and Han, Jing and Hao, Fei},
  journal={Journal of Ambient Intelligence and Humanized Computing},
  volume={14},
  number={6},
  pages={7519--7530},
  year={2023},
  publisher={Springer}
}

@article{belaza2019social,
  title={Social stability and extended social balance—Quantifying the role of inactive links in social networks},
  author={Belaza, Andres M and Ryckebusch, Jan and Bramson, Aaron and Casert, Corneel and Hoefman, Kevin and Schoors, Koen and van den Heuvel, Milan and Vandermarliere, Benjamin},
  journal={Physica A: Statistical Mechanics and its Applications},
  volume={518},
  pages={270--284},
  year={2019},
  publisher={Elsevier}
}

@article{doreian2009partitioning,
  title={Partitioning signed social networks},
  author={Doreian, Patrick and Mrvar, Andrej},
  journal={Social Networks},
  volume={31},
  number={1},
  pages={1--11},
  year={2009},
  publisher={Elsevier}
}

@phdthesis{diaz2025mathematical,
  title={Mathematical analysis of signed networks: structure and dynamics},
  author={Diaz-Diaz, Fernando},
  year={2025},
  school={Institute of Cross-Disciplinary Physics and Complex Systems, IFISC}
}

@article{zigron2019help,
  title={“Help is where you find it”: The role of weak ties networks as sources of information and support in virtual health communities},
  author={Zigron, Shimrit and Bronstein, Jenny},
  journal={Journal of the Association for Information Science and Technology},
  volume={70},
  number={2},
  pages={130--139},
  year={2019},
  publisher={Wiley Online Library}
}

@article{ryberg2008networked,
  title={Networked identities: understanding relationships between strong and weak ties in networked environments},
  author={Ryberg, Thomas and Larsen, Malene Charlotte},
  journal={Journal of Computer Assisted Learning},
  volume={24},
  number={2},
  pages={103--115},
  year={2008},
  publisher={Wiley Online Library}
}

@article{baysha2020dividing,
  title={Dividing social networks: Facebook unfriending, unfollowing, and blocking in turbulent political times},
  author={Baysha, Olga},
  journal={Russian Journal of communication},
  volume={12},
  number={2},
  pages={104--120},
  year={2020},
  publisher={Taylor \& Francis}
}

@article{kaiser2022partisan,
  title={Partisan blocking: Biased responses to shared misinformation contribute to network polarization on social media},
  author={Kaiser, Johannes and Vaccari, Cristian and Chadwick, Andrew},
  journal={Journal of Communication},
  volume={72},
  number={2},
  pages={214--240},
  year={2022},
  publisher={Oxford University Press}
}

@article{zhu2022political,
  title={Political implications of disconnection on social media: A study of politically motivated unfriending},
  author={Zhu, Qinfeng and Skoric, Marko M},
  journal={New Media \& Society},
  volume={24},
  number={12},
  pages={2659--2679},
  year={2022},
  publisher={Sage Publications Sage UK: London, England}
}

@article{seckin2025identifying,
  title={Identifying Constructive Conflict in Online Discussions through Controversial yet Toxicity Resilient Posts},
  author={Seckin, Ozgur Can and Truong, Bao Tran and Flammini, Alessandro and Menczer, Filippo},
  journal={arXiv preprint arXiv:2509.18303},
  year={2025}
}

@article{humprecht2020hostile,
  title={Hostile emotions in news comments: A cross-national analysis of Facebook discussions},
  author={Humprecht, Edda and Hellmueller, Lea and Lischka, Juliane A},
  journal={Social Media+ Society},
  volume={6},
  number={1},
  pages={2056305120912481},
  year={2020},
  publisher={Sage Publications Sage UK: London, England}
}

@article{scheuerman2021framework,
  title={A framework of severity for harmful content online},
  author={Scheuerman, Morgan Klaus and Jiang, Jialun Aaron and Fiesler, Casey and Brubaker, Jed R},
  journal={Proceedings of the ACM on Human-Computer Interaction},
  volume={5},
  number={CSCW2},
  pages={1--33},
  year={2021},
  publisher={ACM New York, NY, USA}
}

@inproceedings{xu2013structures,
  title={Structures of broken ties: exploring unfollow behavior on twitter},
  author={Xu, Bo and Huang, Yun and Kwak, Haewoon and Contractor, Noshir},
  booktitle={Proceedings of the 2013 conference on Computer supported cooperative work},
  pages={871--876},
  year={2013}
}

@inproceedings{saveski2021structure,
  title={The structure of toxic conversations on Twitter},
  author={Saveski, Martin and Roy, Brandon and Roy, Deb},
  booktitle={Proceedings of the web conference 2021},
  pages={1086--1097},
  year={2021}
}

@inproceedings{kumar2023understanding,
  title={Understanding the behaviors of toxic accounts on reddit},
  author={Kumar, Deepak and Hancock, Jeff and Thomas, Kurt and Durumeric, Zakir},
  booktitle={Proceedings of the ACM web conference 2023},
  pages={2797--2807},
  year={2023}
}

@inproceedings{myers2014bursty,
  title={The bursty dynamics of the twitter information network},
  author={Myers, Seth A and Leskovec, Jure},
  booktitle={Proceedings of the 23rd international conference on World wide web},
  pages={913--924},
  year={2014}
}

@incollection{snijders2017modeling,
  title={Modeling the coevolution of networks and behavior},
  author={Snijders, Tom and Steglich, Christian and Schweinberger, Michael},
  booktitle={Longitudinal models in the behavioral and related sciences},
  pages={41--71},
  year={2017},
  publisher={Routledge}
}

@article{goel2023hatemongers,
  title={Hatemongers ride on echo chambers to escalate hate speech diffusion},
  author={Goel, Vasu and Sahnan, Dhruv and Dutta, Subhabrata and Bandhakavi, Anil and Chakraborty, Tanmoy},
  journal={PNAS nexus},
  volume={2},
  number={3},
  pages={pgad041},
  year={2023},
  publisher={Oxford University Press US}
}

@article{israeli2022going,
  title={Going extreme: Comparative analysis of hate speech in parler and gab},
  author={Israeli, Abraham and Tsur, Oren},
  journal={arXiv preprint arXiv:2201.11770},
  year={2022}
}

@article{baumeister2001bad,
  title={Bad is stronger than good},
  author={Baumeister, Roy F and Bratslavsky, Ellen and Finkenauer, Catrin and Vohs, Kathleen D},
  journal={Review of general psychology},
  volume={5},
  number={4},
  pages={323--370},
  year={2001},
  publisher={SAGE Publications Sage CA: Los Angeles, CA}
}

@article{rozin2001negativity,
  title={Negativity bias, negativity dominance, and contagion},
  author={Rozin, Paul and Royzman, Edward B},
  journal={Personality and social psychology review},
  volume={5},
  number={4},
  pages={296--320},
  year={2001},
  publisher={Sage Publications Sage CA: Los Angeles, CA}
}

@inproceedings{kunegis2010spectral,
  title={Spectral analysis of signed graphs for clustering, prediction and visualization},
  author={Kunegis, J{\'e}r{\^o}me and Schmidt, Stephan and Lommatzsch, Andreas and Lerner, J{\"u}rgen and De Luca, Ernesto W and Albayrak, Sahin},
  booktitle={Proceedings of the 2010 SIAM international conference on data mining},
  pages={559--570},
  year={2010},
  organization={SIAM}
}

@article{cartwright1956structural,
  title={Structural balance: a generalization of Heider's theory.},
  author={Cartwright, Dorwin and Harary, Frank},
  journal={Psychological review},
  volume={63},
  number={5},
  pages={277},
  year={1956},
  publisher={American Psychological Association}
}

@article{heider1946attitudes,
  title={Attitudes and cognitive organization},
  author={Heider, Fritz},
  journal={The Journal of psychology},
  volume={21},
  number={1},
  pages={107--112},
  year={1946},
  publisher={Taylor \& Francis}
}

@inproceedings{chiang2011exploiting,
  title={Exploiting longer cycles for link prediction in signed networks},
  author={Chiang, Kai-Yang and Natarajan, Nagarajan and Tewari, Ambuj and Dhillon, Inderjit S},
  booktitle={Proceedings of the 20th ACM international conference on Information and knowledge management},
  pages={1157--1162},
  year={2011}
}

@inproceedings{cheng2025bts,
  title={BTS: A Comprehensive Benchmark for Tie Strength Prediction},
  author={Cheng, Xueqi and Yang, Catherine and Zhao, Yuying and Wang, Yu and Karimi, Hamid and Derr, Tyler},
  booktitle={Proceedings of the 31st ACM SIGKDD Conference on Knowledge Discovery and Data Mining V. 2},
  pages={5345--5354},
  year={2025}
}

@inproceedings{cheng2025edge,
  title={Edge-Centric Network Analytics},
  author={Cheng, Xueqi},
  booktitle={Proceedings of the Eighteenth ACM International Conference on Web Search and Data Mining},
  pages={1071--1073},
  year={2025}
}

@inproceedings{cheng2025edgeclass,
  title={Edge classification on graphs: new directions in topological imbalance},
  author={Cheng, Xueqi and Wang, Yu and Liu, Yunchao and Zhao, Yuying and Aggarwal, Charu C and Derr, Tyler},
  booktitle={Proceedings of the Eighteenth ACM International Conference on Web Search and Data Mining},
  pages={392--400},
  year={2025}
}

\end{document}